\documentclass[
singlecolumn,
pra,showpacs,preprintnumbers,amsmath,amssymb]{revtex4}

\usepackage{graphicx}
\usepackage{amssymb}
\usepackage{amsmath}
\usepackage{dsfont}
\usepackage{bm}
\usepackage{mathrsfs}
\usepackage{times}

\begin{document}
\title{Casimir-Lifshitz force between moving plates at different temperatures}
\author{T.\ G.\ Philbin}
\email{tgp3@st-andrews.ac.uk}
\author{U.\ Leonhardt}
\affiliation{School of Physics and Astronomy, University of St Andrews,
North Haugh, St Andrews, Fife, KY16 9SS, Scotland, UK}

\begin{abstract}
The zero-temperature Casimir-Lifshitz force between two plates moving parallel to each other at arbitrary constant speed was found in [New J.\ Phys.\ {\bf 11}, 033035 (2009)]. The solution is here generalized to the case where the plates are at different temperatures. The Casimir-Lifshitz force is obtained by calculating the electromagnetic stress tensor, using the method employed by Antezza {\it et al.}  [Phys.\ Rev.\ A {\bf 77}, 022901 (2008)] for non-moving plates at different temperatures. The perpendicular force on the plates has contributions from the quantum vacuum and from the thermal radiation; both of these contributions are influenced by the motion. In addition to the perpendicular force, thermal radiation from the moving plates gives rise to a lateral component of the Casimir-Lifshitz force, an effect with no quantum-vacuum contribution. The zero-temperature results are reproduced, in particular the non-existence of a quantum-vacuum friction between the plates.
\hspace*{\fill}
\end{abstract}
\date{\today}

\maketitle

\section{Introduction}
The theory of the electromagnetic force between macroscopic bodies due to vacuum zero-point energy was developed by Lifshitz and co-workers~\cite{lif55,LL,dzy61}, following the consideration by Casimir of perfect mirrors~\cite{cas48}. Recently, Lifshitz's formula~\cite{lif55,LL} for the quantum-vacuum force between two parallel dielectric plates was generalized to the case where the plates have an arbitrary constant lateral motion~\cite{phi09}. The motion introduces considerable extra complexity into the calculation but the final answer turned out to be relatively simple (it will be recovered here as a special case). As described in~\cite{phi09}, a contradictory literature had built up around this problem, with claims of a lateral (``quantum friction") force on the plates but disagreement as to its magnitude. The exact solution~\cite{phi09} showed that, whereas  the perpendicular force is altered by the motion, there is in fact no lateral force on the plates.

Lifshitz~\cite{lif55,LL} also included the effect of thermal radiation in his analysis. The {\it Casimir-Lifshitz effect} is therefore also taken to describe forces that have a contribution from thermal radiation as well as from the quantum vacuum. The formalism developed by Lifshitz, however, cannot be used for plates at different temperatures. The problem of non-moving parallel plates at different temperatures has been comprehensively studied by Antezza {\it et al.}~\cite{ant08}. These authors used Rytov's theory of electromagnetic fluctuations~\cite{rytov} to compute separately the contribution of each plate to the electromagnetic stress tensor, which gives the Casimir-Lifshitz force. The approach is similar to Polder and van Hove's~\cite{pol71} treatment of radiative heat transfer between plates. The Casimir-Lifshitz effect out of thermal equilibrium has been investigated experimentally by measuring the surface-atom force between a Bose-Einstein condensate and a dielectric substrate~\cite{obr07}; the measurement was done for a substrate temperature higher than that of the surrounding environment and results were in agreement with theory~\cite{ant05}.

In this paper we use the method of Antezza {\it et al.}~\cite{ant08} to compute the Casimir-Lifshitz force between plates at different temperatures when one of the plates moves at an arbitrary constant speed parallel to the other. The zero-temperature result~\cite{phi09} therefore emerges as a special case. As a consequence of the thermal radiation, the electromagnetic stress tensor has non-zero off-diagonal components which cause a lateral force on the plates. The lateral force acts against the relative motion and  is present even if the plates have the same (non-zero) temperature. Thus, the thermal radiation provides the means of restoring thermodynamic equilibrium, which requires the damping of the relative motion. The thermal radiation from each plate has a Planck spectrum $1/[\exp(\hbar\omega/k_BT)-1]$ in a frame co-moving with that plate and since the frequency changes under a Lorentz transformation, the spectrum of the radiation from each plate is different in frames with relative motion. This change in the thermal spectrum of the radiation from one plate when viewed in a frame co-moving with the other causes the lateral force between plates at the same temperature. The disappearance of the lateral force at zero temperature can therefore be attributed to the Lorentz invariance of quantum-vacuum energy

\begin{figure}[t]
\begin{center}
\includegraphics[width=20.0pc]{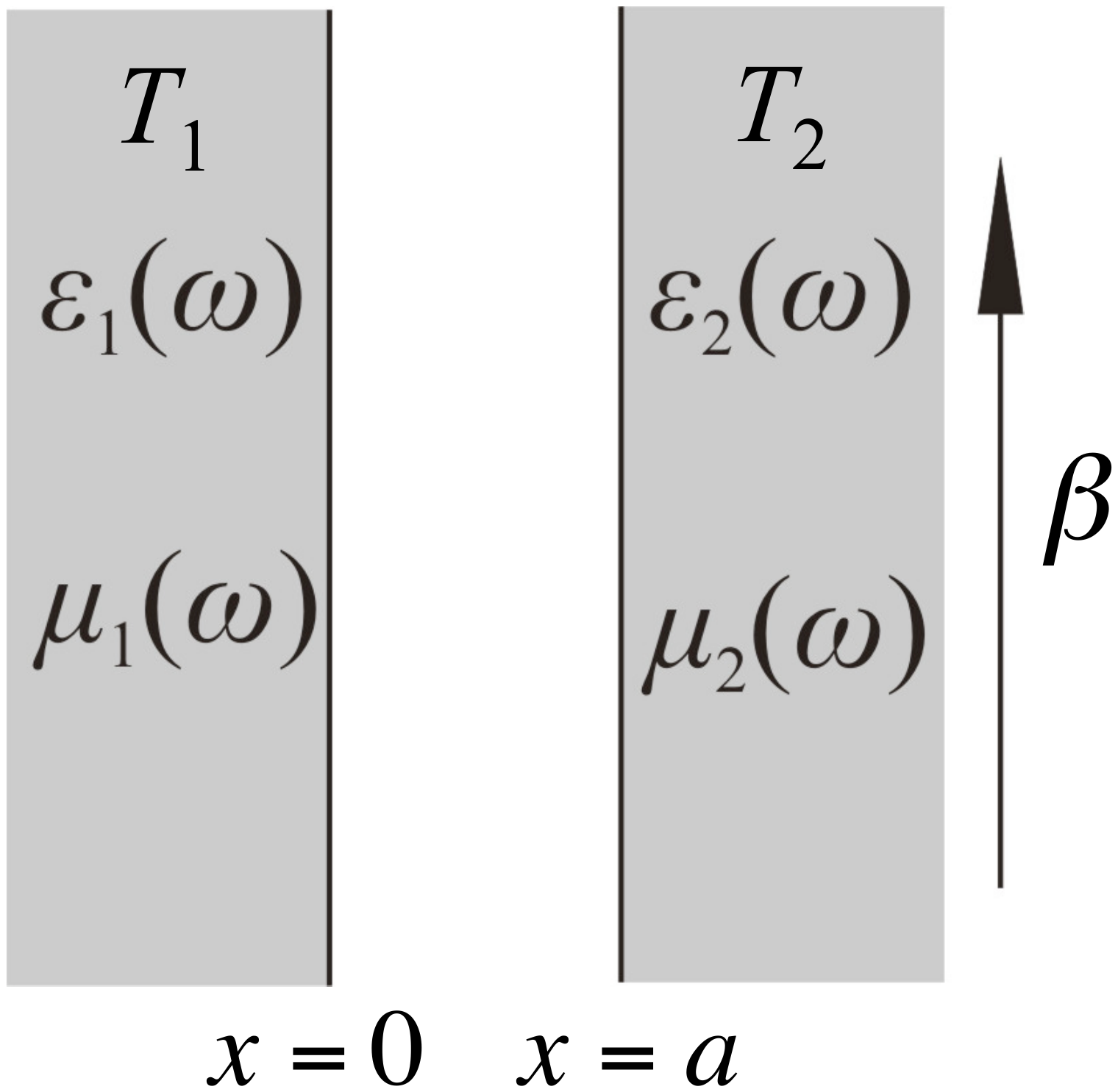}
\vspace*{-10mm}
\end{center}
\caption{Two plates at different temperatures, with general electric permitivities and magnetic permeabilities, lie in the $yz$-plane with constant separation $a$. The second plate moves in the postive $y$-direction with speed $\beta c$, where $c$ is the speed of light. We compute the Casimir-Lifshitz force on Plate~1. \label{fig1}}
\end{figure}

The purpose of this paper is to derive the exact Casimir-Lifshitz force for this problem; the plates are at different temperatures and have arbitrary electric permittivities and magnetic permeabilities, and arbitrary constant lateral motion. We will not here enter into considerations of specific materials or approximations that hold in particular regimes, since the exact solution already requires a lengthy derivation. Moreover, the subject of Casimir-Lifshitz forces on moving media has suffered from a conflict of assertions based on various attempts at approximations (see references and discussion in~\cite{phi09}) and we wish to make as clear as possible that the problem considered here is treated in full generality, within the framework of macroscopic electromagnetism. The geometry and notation of the problem are set out in Figure~\ref{fig1} and caption. The plates have are taken to have a separation $a$ in the $x$-direction. Plate~2 moves in the positive $y$-direction and we use $\beta$ to denote its speed in units of the speed of light $c$, i.e.\ the speed is $\beta c$. The formalism used in the calculation is described in Section~\ref{form}; it is essentially that of Antezza {\it et al.}~\cite{ant08}, except that we allow for a magnetic response in the plates. This results in an expression for the contribution of each plate to the electromagnetic stress tensor in terms of a Green tensor of the vector potential. The contribution of the non-moving plate is found in Section~\ref{plate1}; this involves the computation of the required Green tensor. In Section~\ref{plate2}, the contribution of the moving plate is found from that of the non-moving one by taking account of its differing material properties and its motion. Finally, the Casimir-Lifshitz force is obtained in Section~\ref{CL}.

\section{Formalism} \label{form}
Casimir-Lifshitz forces are given by the expectation value of the electromagnetic stress tensor~\cite{lif55,LL,dzy61}
\begin{equation} \label{stress}
\bm{\sigma}=\langle \mathbf{\hat{D}}\otimes\mathbf{\hat{E}}\rangle+\langle \mathbf{\hat{H}}\otimes\mathbf{\hat{B}}\rangle
-\frac{1}{2}\mathds{1}(\langle \mathbf{\hat{D}}\cdot\mathbf{\hat{E}}\rangle+\langle \mathbf{\hat{H}}\cdot\mathbf{\hat{B}}\rangle).
\end{equation}
For zero-temperature problems the averaging in (\ref{stress}) is simply the (electromagnetic) vacuum expectation value, whereas if thermal effects are included it also contains the thermal averaging. Since the plates are at different temperatures the contribution of each to the fluctuating electromagnetic fields is computed separately. The relevant components of the stress tensor are $\sigma_{xx}$, which gives the perpendicular force on the plates, and $\sigma_{xy}$ ($=\sigma_{yx}$) which gives any non-vanishing lateral force on the plates. We denote the contributions of Plate~1 at temperature $T_1$ to these components by $\sigma_{1xx}(T_1)$ and $\sigma_{1xy}(T_1)$, with a similar notation for the contributions of Plate~2 at temperature $T_2$. Thus, we write the relevant components of the stress tensor as
\begin{gather}
\sigma_{xx}(T_1,T_2)=\sigma_{1xx}(T_1)+\sigma_{2xx}(T_2), \label{xx}  \\
\sigma_{xy}(T_1,T_2)=\sigma_{1xy}(T_1)+\sigma_{2xy}(T_2).  \label{xy}
\end{gather}

We use a gauge in which the scalar potential is zero, so that the electric and magnetic fields are given by
\begin{equation}  \label{EBdef}
\mathbf{\hat{E}}=-\partial_t\mathbf{\hat{A}}, \qquad \mathbf{\hat{B}}=\nabla\times\mathbf{\hat{A}},
\end{equation}
and the vector-potential wave equation is
\begin{eqnarray} 
\mathbf{\hat{A}}(\mathbf{r},\omega)=\int^\infty_{-\infty} d t \,\mathbf{\hat{A}}(\mathbf{r},t)\,e^{i \omega t},  \label{freqA} \\[5pt]
\left(\nabla\times\frac{1}{\mu(\omega)}\nabla\times-\varepsilon(\omega)\frac{\omega^2}{c^2}\right)\mathbf{\hat{A}}=\mu_0\mathbf{\hat{j}}, \label{Awave}
\end{eqnarray}
where $\mathbf{\hat{j}}$ is the current density of any sources. The retarded Green tensor for the vector potential satisfies
\begin{equation}  \label{green}
\left(\nabla\times\frac{1}{\mu(\omega)}\nabla\times-\varepsilon(\omega)\frac{\omega^2}{c^2}\right)\mathbf{G}(\mathbf{r},\mathbf{r'},\omega)=\mathds{1}\delta(\mathbf{r}-\mathbf{r'}).
\end{equation}
and relates the vector potential at $\mathbf{r}$ to the current density at $\mathbf{r'}$:
\begin{equation}  \label{AGj}
\mathbf{\hat{A}}(\mathbf{r},\omega)=\mu_0\int d^3\mathbf{r'}\,\mathbf{G}(\mathbf{r},\mathbf{r'},\omega)\cdot\mathbf{\hat{j}}(\mathbf{r'},\omega).
\end{equation}
In Rytov's formalism~\cite{rytov}, the fluctuating electromagnetic fields are regarded as being produced by fictitious fluctuating currents through (\ref{AGj}). 

The expectation values that occur in the stress tensor (\ref{stress}) can all be calculated from an equal-time correlation function of the vector potential operator.  To avoid ambiguities in the operator ordering we follow the usual quantum prescription and consider the correlation function
\begin{equation}  \label{AA}
\langle \hat{A}_a(\mathbf{r},t)\hat{A}_b(\mathbf{r'},t)\rangle_S
:=\frac{1}{2}\langle \hat{A}_a(\mathbf{r},t)\hat{A}_b(\mathbf{r'},t)+\hat{A}_b(\mathbf{r'},t)\hat{A}_a(\mathbf{r},t)\rangle. 
\end{equation}
Using (\ref{freqA}) and (\ref{AGj}), we can relate (\ref{AA}) to a similar correlation function of the fictitious currents:
\begin{align} 
\langle \hat{A}_a(\mathbf{r},t)\hat{A}_b(\mathbf{r'},t)\rangle_S=&\frac{\mu_0^2}{(2\pi)^2}\int^\infty_{-\infty} d\omega\int^\infty_{-\infty} d\omega'\int d^3\mathbf{r''}\int d^3\mathbf{r'''}\,G_{ac}(\mathbf{r},\mathbf{r''},\omega)G_{bd}(\mathbf{r'},\mathbf{r'''},\omega')  \nonumber\\
&\times\langle \hat{j}_c(\mathbf{r''},\omega)\hat{j}_d(\mathbf{r'''},\omega')\rangle_S \,e^{-i(\omega+\omega')t}.   \label{AAjj}
\end{align}
The correlation function of the currents is given by the fluctuation-dissipation theorem~\cite{LLSP1,LL} as
\begin{align}
\langle \hat{j}_c(\mathbf{r''},\omega)\hat{j}_d(\mathbf{r'''},\omega')\rangle_S=&\frac{2\pi\hbar}{\mu_0}\left[\frac{\omega^2}{c^2}\text{Im}(\varepsilon(\omega))\delta_{cd}+\frac{\text{Im}(\mu(\omega))}{|\mu(\omega)|^2}\left(\nabla''\cdot\nabla'''\delta_{cd}-\nabla''_c\nabla'''_d\right)\right]\delta(\mathbf{r''}-\mathbf{r'''}) \nonumber \\
&\times\delta(\omega+\omega')\coth \left(\frac{\hbar\omega}{2k_BT}\right). \label{jj}
\end{align}
The right-hand side of (\ref{jj}) is constructed from the differential operator in (\ref{green}) that functions as the ``inverse" of the Green tensor, which in turn serves as a generalized susceptibility~\cite{LLSP1,LL} for the vector potential through (\ref{AGj}). The derivatives in  (\ref{jj})  act on the delta function $\delta(\mathbf{r''}-\mathbf{r'''})$ and have been written in a form that makes clearer the symmetry under interchange of $\mathbf{r''}$ and $\mathbf{r'''}$; throughout this paper we use Cartesian coordinates, for which $\nabla''_a\delta(\mathbf{r''}-\mathbf{r'''})=-\nabla'''_a\delta(\mathbf{r''}-\mathbf{r'''})$. Equation~(\ref{jj}) is usually written without the term containing the magnetic permeability, as in~\cite{ant08} for example. Since we allow for arbitrary magnetic permeabilities in the plates we must write the more cumbersome expression (\ref{jj}). The final formulae for the contributions to the electromagnetic stress tensor will all be written in terms of reflection coefficients for transverse electric (TE) and transverse magnetic (TM) plane waves, and the permittivities and permeabilities will appear only through these reflection coefficients. Not surprisingly, the same result is obtained if one ignores the magnetic response in the derivation but then includes it in the reflections coefficients once the final expression is obtained.

Using (\ref{jj}) and the Green tensor property~\cite{LL,phi09} 
\begin{equation}  \label{Gprop}
G_{bd}(\mathbf{r'},\mathbf{r'''},-\omega)=G^*_{bd}(\mathbf{r'},\mathbf{r'''},\omega) \qquad \text{($\omega$ real)}
\end{equation}
we can write (\ref{AAjj}) as
\begin{equation} \label{AAfreq1}
\langle \hat{A}_a(\mathbf{r},t)\hat{A}_b(\mathbf{r'},t)\rangle_S=\int^\infty_{-\infty} d\omega\,\langle \mathbf{\hat{A}}\mathbf{\hat{A}}\rangle_{ab}(\mathbf{r},\mathbf{r'},\omega),
\end{equation}
where $\langle \mathbf{\hat{A}}\mathbf{\hat{A}}\rangle_{ab}(\mathbf{r},\mathbf{r'},\omega)$ is a frequency-space correlation function defined by
\begin{align} 
\langle \mathbf{\hat{A}}\mathbf{\hat{A}}\rangle_{ab}(\mathbf{r},\mathbf{r'},\omega)=&\frac{\hbar\mu_0}{2\pi}\int d^3\mathbf{r''}\int d^3\mathbf{r'''}\,\coth \left(\frac{\hbar\omega}{2k_BT}\right)G_{ac}(\mathbf{r},\mathbf{r''},\omega)G^*_{bd}(\mathbf{r'},\mathbf{r'''},\omega)  \nonumber\\[5pt]
&  \times\left[\frac{\omega^2}{c^2}\text{Im}(\varepsilon(\omega))\delta_{cd}+\frac{\text{Im}(\mu(\omega))}{|\mu(\omega)|^2}\left(\nabla''\cdot\nabla'''\delta_{cd}-\nabla''_c\nabla'''_d\right)\right]\delta(\mathbf{r''}-\mathbf{r'''}). \label{AAGG}
\end{align}
(We note in passing that the familiar property of the Green tensor for non-moving media, namely that $G_{bd}(\mathbf{r'},\mathbf{r'''},\omega)$ is equal to $G_{db}(\mathbf{r'''},\mathbf{r'},\omega)$, does \textit{not} hold for the problem considered here since it requires time-reversal invariance~\cite{LL}.) Equation (\ref{Gprop}) and the similar property of the permittivity and permeability~\cite{LLcm} allows us replace $\int^\infty_{-\infty}d\omega$ by $2\text{Re}\int^\infty_{0}d\omega$ in (\ref{AAfreq1}):
\begin{equation} \label{AAfreq}
\langle \hat{A}_a(\mathbf{r},t)\hat{A}_b(\mathbf{r'},t)\rangle_S=2\text{Re}\int^\infty_{0}d\omega\,\langle \mathbf{\hat{A}}\mathbf{\hat{A}}\rangle_{ab}(\mathbf{r},\mathbf{r'},\omega),
\end{equation}
Correlation functions for the electric and magnetic fields are easily obtained by using (\ref{EBdef}) and (\ref{freqA}) in the foregoing derivation, with the result:
\begin{align} 
\langle \hat{E}_a(\mathbf{r},t)\hat{E}_b(\mathbf{r'},t)\rangle_S&=2\text{Re}\int^\infty_{0}d\omega\,\omega^2\langle \mathbf{\hat{A}}\mathbf{\hat{A}}\rangle_{ab}(\mathbf{r},\mathbf{r'},\omega),  \label{EE}  \\
\langle \hat{B}_a(\mathbf{r},t)\hat{B}_b(\mathbf{r'},t)\rangle_S&=2\text{Re}\int^\infty_{0}d\omega\,\epsilon_{acd}\epsilon_{bef}\nabla_c\nabla'_e\langle \mathbf{\hat{A}}\mathbf{\hat{A}}\rangle_{df}(\mathbf{r},\mathbf{r'},\omega),  \label{BB}
\end{align}
where $\epsilon_{acd}$ is (in Cartesian coordinates) the permutation symbol.  Equations (\ref{EE})--(\ref{BB}) are suitable for finding the stress tensor (\ref{stress}) between the plates, where $\mathbf{D}=\varepsilon_0\mathbf{E}$ and $\mathbf{H}=\mathbf{B}/\mu_0$.

The calculation proceeds by using (\ref{AAGG}) to calculate the contribution of each plate to the correlations functions (\ref{EE})--(\ref{BB}) between the plates, and thence to the stress tensor components (\ref{xx}) and (\ref{xy}). Everything is determined once we find the relevant Green tensor in  (\ref{AAGG}). We first consider the contribution of Plate~1, for which the source points $\mathbf{r''}$, $\mathbf{r'''}$ in  (\ref{AAGG}) are located inside Plate~1 while the points  $\mathbf{r}$, $\mathbf{r'}$ are placed between the plates.

\section{Contribution of Plate~1} \label{plate1}
\subsection{Green tensor}
The required Green tensor can be written down using a method previously employed to solve the zero-temperature case~\cite{phi09}, and also to find the Casimir-Lifshitz forces between birefringent plates~\cite{phi08}. The solution of the mono\-chromatic equation (\ref{green}) has a simple physical meaning: an oscillating dipole at the point $\mathbf{r'}$ emits electromagnetic waves of frequency $\omega$ and $\mathbf{G}(\mathbf{r},\mathbf{r'},\omega)$ is the resulting vector potential at the point $\mathbf{r}$. The second index in $G_{ab}$ represents the orientation of the dipole at $\mathbf{r'}$, while the first index represents the components of the vector potential at $\mathbf{r}$. Here we require the Green tensor  $\mathbf{G}(\mathbf{r},\mathbf{r''},\omega)$ in (\ref{AAGG}), where the dipole sits at  $\mathbf{r''}$ inside Plate~1 and the resulting vector potential is measured at  $\mathbf{r}$ between the plates. To write down the solution we exploit the homogeneity of the problem in the $y$- and $z$- directions to Fourier transform the Green tensor as follows:
\begin{equation} \label{Gfour}
\mathbf{\widetilde{G}}(x,x'',u,v,\omega) 
=\int_{-\infty}^\infty d y\int_{-\infty}^\infty d z\,\mathbf{G}(\mathbf{r},\mathbf{r''},\omega)\,e^{-i u(y-y'')-i v(z-z'')}
\end{equation}
so that we decompose the waves emitted by the dipole into plane waves.  In the absence of the plate boundary, the solution of (\ref{green}) inside the material of Plate~1 is the bare Green tensor~\cite{LL} (with source point labeled $\mathbf{r''}$)
\begin{gather}
\mathbf{\widetilde{G}}_b(x,x'',u,v,\omega)=
\left\{\begin{array}{l} e^{iw(x-x'')}\boldsymbol{\mathcal{G}_+}, \quad x>x''  \\[5pt] 
e^{-iw(x-x'')}\boldsymbol{\mathcal{G}_-}, \quad x<x'' \end{array}\right.,  \label{bare} \\[6pt]
\boldsymbol{\mathcal{G}_\pm}=\frac{c^2}{2\varepsilon_1(\omega)\omega^2w_1}\left(\mathbf{k}_1^\pm \otimes \mathbf{k}_1^\pm+\varepsilon_1(\omega)\mu_1(\omega)\frac{\omega^2}{c^2}\mathds{1}\right),  \label{Gpm}  \\[6pt]
\mathbf{k}_1^\pm=(\pm w_1,u,v), \qquad w_1=\sqrt{\varepsilon_1(\omega)\mu_1(\omega)\frac{\omega^2}{c^2}-s^2}, \qquad s=\sqrt{u^2+v^2}. \label{w1}
\end{gather}
The two possibilities in (\ref{bare}) represent plane waves propagating to the right (first line) or to the left (second line), with wave vectors $\mathbf{k}^\pm$. When $x''$ is in Plate~1 and $x$ is in the gap between the plates, only the right-moving plane wave can reach $x$ from $x''$ and it does this by first passing though the plate boundary into the gap. Once in the gap the plane wave can propagate to $x$ directly, or after an even or odd number of reflections off the plates, giving right- and left-moving plane waves in the gap with wave vectors
\begin{equation}  \label{kwdef}
\mathbf{k}^\pm=(\pm w,u,v), \qquad w=\sqrt{\frac{\omega^2}{c^2}-s^2}.
 \end{equation}
The solution for the Green tensor is the linear superposition of all these possibilities; by the same reasoning employed in~\cite{phi09} and ~\cite{phi08} we can write it as
\begin{equation} \label{Gsol}
\mathbf{\widetilde{G}}(x,x'',u,v,\omega) =\left(e^{iwx}+e^{-iw(x-a)}e^{iwa}\mathbf{R}_2\right)\left(\mathds{1}-e^{2iwa}\mathbf{R}_1\mathbf{R}_2\right)^{-1}e^{iw_1(-x'')}\mathbf{T}\boldsymbol{\mathcal{G}_+}.
\end{equation}
In (\ref{Gsol}), $\mathbf{T}$ is the transition operator (matrix) for the right-moving plane wave in Plate~1 as it passes into the gap, $\mathbf{R}_2$ is the reflection operator for a right-moving plane wave at plate~2 and $\mathbf{R}_1$ is the reflection operator for a left-moving plane wave at plate~1. The inverse matrix in (\ref{Gsol}) is a geometric series representing every possible number (including zero) of double reflections off both plates, the exponentials providing the propagation distance $2a$ for each double reflection. Each term in (\ref{Gsol}), after the series expansion of the inverse matrix, has  an overall exponential factor that accounts for the propagation distance involved. 

The operators $\mathbf{T}$, $\mathbf{R}_2$ and $\mathbf{R}_1$ were derived in~\cite{phi09} by decomposing the plane waves into TE and TM polarizations in a frame co-moving with the relevant plate; they can be written
\begin{gather}
\mathbf{T}=t_{E}\mathbf{n}_{E1}\otimes\mathbf{n}_{E1}-t_{B}\mathbf{n}_{B1}^+\otimes\mathbf{n}_{Bm1},   \label{Tdef}  \\
\mathbf{R}_1=r_{E1}\mathbf{n}_{E1}\otimes\mathbf{n}_{E1}+r_{B1}\mathbf{n}_{B1}^+\otimes\mathbf{n}_{B1}^-,  \qquad \mathbf{R}_2=r_{E2}\mathbf{n}_{E2}^-\otimes\mathbf{n}_{E2}^++r_{B2}\mathbf{n}_{B2}^-\otimes\mathbf{n}_{B2}^+. \label{Rdef}
\end{gather}
The scalar coefficients in (\ref{Tdef}) and (\ref{Rdef}) are transmission and reflection coefficients for TE ($E$) and TM ($B$) polarized plane waves~\cite{jac} evaluated in a frame co-moving with the relevant plate~\cite{phi09}:
\begin{gather}
  t_{E}=\frac{2w_1}{w_1+\mu_1(\omega)w}=1-r_{E1}, \qquad t_{B}=\frac{2\sqrt{\varepsilon_1(\omega)\mu_1(\omega)}w_1}{\mu_1(\omega)w_1+\varepsilon_1(\omega)\mu_1(\omega)w}=\sqrt{\frac{\varepsilon_1(\omega)}{\mu_1(\omega)}}(1+r_{B1}), \label{tdef}  \\
  r_{E1}=\frac{\mu_1(\omega)w-w_1}{\mu_1(\omega)w+w_1}, \quad r_{B1}=-\frac{\varepsilon_1(\omega)w-w_1}{\varepsilon_1(\omega)w+w_1}, \label{r1def} \\[5pt]
  r_{E2}=\frac{\mu_2(\omega')w-w_2}{\mu_2(\omega')w+w_2}, \quad r_{B2}=-\frac{\varepsilon_2(\omega')w-w_2}{\varepsilon_2(\omega')w+w_2},  \quad  w_2=\sqrt{\varepsilon_2(\omega')\mu_2(\omega')\frac{\omega'^2}{c^2}-u'^2-v^2}, \label{r2def}\\[5pt]
 \omega'=\gamma(\omega-\beta c u), \quad u'=\gamma(u-\beta\omega/c), \quad \gamma=(1-\beta^2)^{-\frac{1}{2}}. \label{lor}
\end{gather}
Note that the reflection coefficients (\ref{r2def}) at Plate~2 are those in a frame co-moving with that plate; hence the Lorentz-transformed frequency $\omega'$ and $y$-component of the wave vector $u'$. The unit vectors in in (\ref{Tdef}) and (\ref{Rdef}) are given by~\cite{phi09}
\begin{gather}
\mathbf{n}_{E1}=\frac{1}{s}\left(\begin{array}{c} 0 \\ -v \\ u \end{array}\right),  \qquad
\mathbf{n}_{B1}^\pm=\frac{c}{\omega s}\left(\begin{array}{c} \pm s^2 \\ -uw \\ -vw \end{array}\right),  \qquad \mathbf{n}_{Bm1}=\frac{c}{\sqrt{\varepsilon_1(\omega)\mu_1(\omega)}\,\omega s}\left(\begin{array}{c} -s^2 \\ uw_1 \\ vw_1 \end{array}\right), \\
\mathbf{n}_{E2}^\pm=\frac{c}{\omega\sqrt{s^2-2\beta\omega u/c-\beta^2(v^2-\omega^2/c^2)}} 
\left(\begin{array}{c} \pm\beta vw \\ v(-\omega/c+\beta u) \\ \omega u/c+ \beta(v^2-\omega^2/c^2) \end{array}\right),   \\[5pt]
\mathbf{n}_{B2}^\pm=\frac{c}{\omega\sqrt{s^2-2\beta\omega u/c-\beta^2(v^2-\omega^2/c^2)}} 
\left(\begin{array}{c} \pm(-s^2+\beta\omega u/c) \\ w(u-\beta\omega/c) \\ wv \end{array}\right), \label{nsf}
\end{gather}
and are polarization directions for plane waves, where the superscript $+$ ($-$) refers to a plane wave moving to the right (left).
The Green tensor (\ref{Gsol}) is now completely specified, but before using it to compute the correlation function (\ref{AAGG}) we simplify the inverse matrix in  (\ref{Gsol}). This is done in the Appendix, using the procedure employed in~\cite{phi09}.

\subsection{Contribution to the stress tensor}
In (\ref{Gfour}) and  (\ref{Gsol}) one can see the entire spatial dependence of the Green tensor; spatial derivatives of the Green tensor result in factors of wave-vector components. We insert the fourier-transformed Green tensor (\ref{Gfour}) in (\ref{AAGG}) and integrate by parts so that the derivatives in  (\ref{AAGG}) act on the Green tensors; these derivatives of the Green tensor are evaluated by multiplying by the relevant wave-vector component. The integrations with respect to $\mathbf{r''}$ and $\mathbf{r'''}$ are over the volume of Plate~1; we drop the boundary term at the surface of the plate that results from the integrations by parts, since we can take the fictitious currents to vanish on the surface without affecting the contribution of the plate. This results in the following formula for the correlation function (\ref{AAGG}):
\begin{align} 
\langle \mathbf{\hat{A}}&\mathbf{\hat{A}}\rangle_{ab}(\mathbf{r},\mathbf{r'},\omega)  \nonumber \\
&=\frac{\hbar\mu_0}{4\pi^3}\int^0_{-\infty}dx''\int^\infty_{-\infty}du\int^\infty_{-\infty}dv\,\coth \left(\frac{\hbar\omega}{2k_BT_1}\right) \,e^{i u(y-y')+i v(z-z')} \nonumber\\[5pt]
&\times \widetilde{G}_{ac}(x,x'',u,v,\omega)\widetilde{G}^*_{bd}(x',x'',u,v,\omega) 
 \left\{\frac{\omega^2}{c^2}\text{Im}(\varepsilon_1(\omega))\delta_{cd}+\frac{\text{Im}(\mu_1(\omega))}{|\mu(\omega)|^2}\left[(|w_1|^2+s^2)\delta_{cd}  \right.  \right.  \nonumber \\
&\left. \left. -|w_1|^2\delta_{dx}\delta_{cx}  -u^2\delta_{dy}\delta_{cy}-v^2\delta_{dz}\delta_{cz}-w_1u\delta_{dx}\delta_{cy}-w_1^*u\delta_{dy}\delta_{cx}-w_1v\delta_{dx}\delta_{cz}-w_1^*v\delta_{dz}\delta_{cx}   \right.  \right.  \nonumber \\
& \left. -uv(\delta_{dy}\delta_{cz}+\delta_{dz}\delta_{cy})  \right] \bigg\}. \label{AAGGsim}
\end{align}

The expression for the Green tensor given in the Appendix is now used in (\ref{AAGGsim}) to compute the correlation function. As is clear from the simple dependence of the Green tensor on $x''$, the integration with respect to $x''$ in (\ref{AAGGsim}) is performed using
\begin{equation}
\int^0_{-\infty}dx''\left|e^{-iw_1x''}\right|^2=\frac{1}{2\text{Im}(w_1)}.
\end{equation}
Electric and magnetic field correlation functions follow from (\ref{EE}) and (\ref{BB}), and with $\mathbf{r}'\rightarrow\mathbf{r}$ these determine the contribution to the stress tensor (\ref{stress}) (where $\mathbf{D}=\varepsilon_0\mathbf{E}$ and $\mathbf{H}=\mathbf{B}/\mu_0$ between the plates). The fact that the real part is taken in  (\ref{EE}) and (\ref{BB}) means that the result can be split into terms from evanescent waves ($\omega<cs$) with imaginary $w$ and propagating waves ($\omega>cs$) with real $w$ (see (\ref{kwdef})). Use of the identities
\begin{gather}
\text{Im}(\varepsilon_1)=2\text{Re}(\mu_1)\text{Re}(w_1)\text{Im}(w_1)-\text{Im}(\mu_1)\left\{\frac{c^2}{\omega^2|\mu_1|^2}\left[(\text{Re}(w_1))^2-(\text{Im}(w_1))^2\right]+s^2\right\},  \label{id1} \\[5pt]
\text{Re}(\varepsilon^*_1\mu^*_1w_1)=\frac{c^2}{\omega^2}\text{Re}(w_1)\left(|w_1|^2+s^2\right),    \label{id2}   \\[5pt]
\text{Im}(\varepsilon^*_1\mu^*_1w_1)=-\frac{c^2}{\omega^2}\text{Im}(w_1)\left(|w_1|^2-s^2\right),   \label{id3}
\end{gather}
allows the explicit occurrence of $w_1$, $\varepsilon_1$ and $\mu_1$ to be re-expressed in terms of the reflection coefficients. With the definitions
\begin{gather}
a_{EE}:=1-e^{2iaw}r_{E1}r_{E2}, \quad a_{BB}:=1-e^{2iaw}r_{B1}r_{B2}, \label{adef1} \\
a_{EB}:=1-e^{2iaw}r_{E1}r_{B2}, \quad a_{BE}:=1-e^{2iaw}r_{B1}r_{E2},  \label{adef2}
\end{gather}
the result for $\sigma_{1xx}(T_1)$ and $\sigma_{1xy}(T_1)$ can be written
\begin{align}
\sigma_{1xx}&(T_1)  \nonumber \\
=&\frac{\hbar}{16\pi^3}\int^\infty_{-\infty}du\int^\infty_{-\infty}dv\int^{cs}_{0} d\omega\,\coth \left(\frac{\hbar\omega}{2k_BT_1}\right)e^{-2a|w|}|w|\left[(cs^2-u\beta\omega)^2+v^2\beta^2c^2w^2\right]  \nonumber \\[5pt]
&\times\left\{4\,\text{Im}(r_{E1})\frac{\text{Re}(r_{E2})|a_{BB}|^2(cs^2-u\beta\omega)^2+\text{Re}(r_{B2})|a_{BE}|^2v^2\beta^2c^2w^2}{\left|(cs^2-u\beta\omega)^2a_{EE}a_{BB}+a_{EB}a_{BE}v^2\beta^2c^2w^2\right|^2}+( {\scriptstyle E}\leftrightarrow  {\scriptstyle B})\right\}  \nonumber \\[5pt]
&-\frac{\hbar}{16\pi^3}\int^\infty_{-\infty}du\int^\infty_{-\infty}dv\int^{\infty}_{cs} d\omega\,\coth \left(\frac{\hbar\omega}{2k_BT_1}\right)w\left[(cs^2-u\beta\omega)^2+v^2\beta^2c^2w^2\right]  \nonumber \\[5pt]
&\times\left\{\left(1-\left|r_{E1}\right|^2\right)\frac{\left(1+\left|r_{E2}\right|^2\right)|a_{BB}|^2(cs^2-u\beta\omega)^2+\left(1+\left|r_{B2}\right|^2\right)|a_{BE}|^2v^2\beta^2c^2w^2}{\left|(cs^2-u\beta\omega)^2a_{EE}a_{BB}+a_{EB}a_{BE}v^2\beta^2c^2w^2\right|^2}\right. \nonumber  \\[5pt]
&\ \ \ \ \ \ \ +({\scriptstyle E}\leftrightarrow {\scriptstyle B})\Bigg\},  \label{1xx} 
\end{align}
\begin{align}
\sigma_{1xy}&(T_1)  \nonumber \\
=&-\frac{\hbar}{16\pi^3}\int^\infty_{-\infty}du\int^\infty_{-\infty}dv\int^{cs}_{0} d\omega\,\coth \left(\frac{\hbar\omega}{2k_BT_1}\right)e^{-2a|w|}u\left[(cs^2-u\beta\omega)^2+v^2\beta^2c^2w^2\right]  \nonumber \\[5pt]
&\times\left\{4\,\text{Im}(r_{E1})\frac{\text{Im}(r_{E2})|a_{BB}|^2(cs^2-u\beta\omega)^2+\text{Im}(r_{B2})|a_{BE}|^2v^2\beta^2c^2w^2}{\left|(cs^2-u\beta\omega)^2a_{EE}a_{BB}+a_{EB}a_{BE}v^2\beta^2c^2w^2\right|^2}+( {\scriptstyle E}\leftrightarrow  {\scriptstyle B})\right\}  \nonumber \\[5pt]
&-\frac{\hbar}{16\pi^3}\int^\infty_{-\infty}du\int^\infty_{-\infty}dv\int^{\infty}_{cs} d\omega\,\coth \left(\frac{\hbar\omega}{2k_BT_1}\right)u\left[(cs^2-u\beta\omega)^2+v^2\beta^2c^2w^2\right]  \nonumber \\[5pt]
&\times\left\{\left(1-\left|r_{E1}\right|^2\right)\frac{\left(1-\left|r_{E2}\right|^2\right)|a_{BB}|^2(cs^2-u\beta\omega)^2+\left(1-\left|r_{B2}\right|^2\right)|a_{BE}|^2v^2\beta^2c^2w^2}{\left|(cs^2-u\beta\omega)^2a_{EE}a_{BB}+a_{EB}a_{BE}v^2\beta^2c^2w^2\right|^2}\right. \nonumber  \\[5pt]
&\ \ \ \ \ \ \ +({\scriptstyle E}\leftrightarrow {\scriptstyle B})\Bigg\}.  \label{1xy} 
\end{align}
The first integrals  (\ref{1xx}) and  (\ref{1xy}) come from the evanescent waves while the second integrals are from the propagating waves. It is very important to note that, in addition to their explicit appearance in  (\ref{1xx}) and  (\ref{1xy}), the reflection coefficients at each plate are present in the quantities (\ref{adef1})--(\ref{adef2}).

It is clear on symmetry grounds that the contribution (\ref{1xx}) to the perpendicular force on the plates cannot depend on the sign of $\beta$. One easily sees this to be the case in  (\ref{1xx}) by considering a Taylor expansion of the integrands in powers of $\beta$. Odd powers of $\beta$ are accompanied by odd powers of $u$ and so these terms vanish after the integration with respect to $u$; only the even powers of $\beta$ in the expansion, which are accompanied by even powers of $u$, contribute to $\sigma_{1xx}(T_1)$. (Note that the reflection coefficients (\ref{r2def})--(\ref{lor}) at Plate~2 must also be expanded in powers of $\beta$ to obtain the Taylor expansion of the integrands.) Similar symmetry considerations show that the contribution (\ref{1xy}) to the lateral force must change sign when $\beta$ changes sign. This is also seen to be the case by a Taylor expansion of the integrands. Because of the overall factor of $u$ in the integrands in (\ref{1xy}) it is the terms with odd powers of $\beta$ that are now accompanied by an even power of $u$ and so contribute to  $\sigma_{1xy}(T_1)$, whereas the terms even in $\beta$ vanish after the integration with respect to $u$. When $\beta=0$, $\sigma_{1xy}(T_1)$ vanishes and $\sigma_{1xx}(T_1)$ reduces to the result of Antezza {\it et al.}~\cite{ant08} for non-moving plates, as one can see by inspection.

\subsection{Contribution to the Poynting vector}
In the next section we will find that the contribution to the Poynting vector between the plates enters into our method of calculating the Casimir-Lifshitz force. The Poynting vector between the plates
\begin{equation} \label{poyn}
\mathbf{S}=\frac{1}{\mu_0}\langle\mathbf{\hat{E}}\times\mathbf{\hat{B}}\rangle
\end{equation}
can also be obtained from the correlation function (\ref{AAGGsim}). Similar to (\ref{EE}) and (\ref{BB}), we obtain an electric-magnetic correlation function given by
\begin{equation}  \label{EB}
\langle \hat{E}_a(\mathbf{r},t)\hat{B}_b(\mathbf{r'},t)\rangle_S=2\text{Re}\,\frac{i}{\mu_0}\int^\infty_{0}d\omega\,\omega\,\epsilon_{bcd}\nabla'_c\langle \mathbf{\hat{A}}\mathbf{\hat{A}}\rangle_{ad}(\mathbf{r},\mathbf{r'},\omega).
\end{equation}
The object of interest is the contribution of Plate~ 1 to the $x$-component of the Poynting vector, which we denote by $S_{1x}(T_1)$. From (\ref{poyn}) and (\ref{EB}) we see that $S_{1x}(T_1)$ is determined by (\ref{AAGGsim}) as
\begin{equation} \label{poynx}
S_{1x}(T_1)=2\text{Re}\,\frac{i}{\mu_0}\int^\infty_{0}d\omega\,\omega\,\epsilon_{xab}\epsilon_{bcd}\left.\nabla'_c\langle \mathbf{\hat{A}}\mathbf{\hat{A}}\rangle_{ad}(\mathbf{r},\mathbf{r'},\omega)\right|_{\mathbf{r'}=\mathbf{r}}.
\end{equation}
Inserting (\ref{AAGGsim}) in (\ref{poynx}), and again making use of the identities (\ref{id1})--(\ref{id3}) and the definitions (\ref{adef1})--(\ref{adef2}), we obtain
\begin{align}
S_{1x}&(T_1)  \nonumber \\
=&\frac{\hbar}{16\pi^3}\int^\infty_{-\infty}du\int^\infty_{-\infty}dv\int^{cs}_{0} d\omega\,\coth \left(\frac{\hbar\omega}{2k_BT_1}\right)e^{-2a|w|}\omega\left[(cs^2-u\beta\omega)^2+v^2\beta^2c^2w^2\right]  \nonumber \\[5pt]
&\times\left\{4\,\text{Im}(r_{E1})\frac{\text{Im}(r_{E2})|a_{BB}|^2(cs^2-u\beta\omega)^2+\text{Im}(r_{B2})|a_{BE}|^2v^2\beta^2c^2w^2}{\left|(cs^2-u\beta\omega)^2a_{EE}a_{BB}+a_{EB}a_{BE}v^2\beta^2c^2w^2\right|^2}+( {\scriptstyle E}\leftrightarrow  {\scriptstyle B})\right\}  \nonumber \\[5pt]
&\frac{\hbar}{16\pi^3}\int^\infty_{-\infty}du\int^\infty_{-\infty}dv\int^{\infty}_{cs} d\omega\,\coth \left(\frac{\hbar\omega}{2k_BT_1}\right)\omega\left[(cs^2-u\beta\omega)^2+v^2\beta^2c^2w^2\right]  \nonumber \\[5pt]
&\times\left\{\left(1-\left|r_{E1}\right|^2\right)\frac{\left(1-\left|r_{E2}\right|^2\right)|a_{BB}|^2(cs^2-u\beta\omega)^2+\left(1-\left|r_{B2}\right|^2\right)|a_{BE}|^2v^2\beta^2c^2w^2}{\left|(cs^2-u\beta\omega)^2a_{EE}a_{BB}+a_{EB}a_{BE}v^2\beta^2c^2w^2\right|^2}\right. \nonumber  \\[5pt]
&\ \ \ \ \ \ \ +({\scriptstyle E}\leftrightarrow {\scriptstyle B})\Bigg\}.  \label{S1x} 
\end{align}
Note that the integrands in (\ref{S1x}) differ from those in (\ref{1xy}) only by a factor of $-\frac{\omega}{u}$, a fact that will have great significance for the Casimir-Lifshitz force.

\section{Contribution of Plate~2} \label{plate2}
To obtain the contribution of Plate~2 to the Casimir-Lifshitz force we use the results of the previous section to write down its contribution to the stress tensor and the Poynting vector in a frame co-moving with Plate~2. In this frame the radiating sources in Plate~2 are at rest and Plate~1 is moving in the negative $y$-direction (speed $-\beta$ in the $y$-direction). Once we have obtained the contributions of Plate~2 in the co-moving frame we will consider how they determine the contributions $\sigma_{2xx}(T_2)$ and $\sigma_{2xy}(T_2)$ to the stress tensor in the frame of Fig.~\ref{fig1}. 

\subsection{Contribution in co-moving frame}
Consider the changes needed to turn the results  (\ref{1xx}),  (\ref{1xy}) and (\ref{S1x}) for  $\sigma_{1xx}(T_1)$,  $\sigma_{1xy}(T_1)$ and $S_{1x}(T_1)$ into  $\sigma'_{2xx}(T_2)$,  $\sigma'_{2xy}(T_2)$ and $S'_{2x}(T_2)$, where the prime refers to a frame co-moving with Plate~2. Clearly, the interchange $1\leftrightarrow 2$ is required, as regards the reflection coefficients and temperature. Note that the reflection coefficients (\ref{r1def})-- (\ref{r2def}) are evaluated in frames co-moving with each plate so we use the same reflection coefficients as in the frame of Fig.~\ref{fig1}. Thus the only change required in the reflection coefficients and temperature in  (\ref{1xx}),  (\ref{1xy}) and (\ref{S1x}) is the interchange $1\leftrightarrow 2$. The radiating plate (Plate~2) is now at $x=a$ with the non-radiating plate at $x=0$, so compared to the last section this entails the following change in the $x$-coordinate:
\begin{equation} \label{xtrans}
x\rightarrow -x +a.
\end{equation}
Neither the stress tensor contributions (\ref{1xx}) and (\ref{1xy}), nor the Poynting vector contribution (\ref{S1x}), are functions of $x$, but the transformation (\ref{xtrans}) affects the tensor and vector components~\cite{Telephone}:
\begin{gather}
\Phi_{ab}=
\left(
\begin{array}{ccc}
-1  & 0  & 0  \\
0  & 1  & 0  \\
0  & 0  & 1  
\end{array}
\right),  \\
\sigma_{xx}\rightarrow\Phi_{xa}\Phi_{xb}\,\sigma_{ab}=\sigma_{xx}, \qquad \sigma_{xy}\rightarrow\Phi_{xa}\Phi_{yb}\,\sigma_{1ab}=-\sigma_{xy}, \\
S_x\rightarrow\Phi_{xa}S_a=-S_x.
\end{gather}
Thus, in addition to the other changes, a sign change is required in (\ref{1xy}) and (\ref{S1x}) to obtain the contribution of Plate~2, whereas this is not the case for (\ref{1xx}). The moving plate (Plate~1) now has speed $-\beta$ and the integration variables  (\ref{1xx}) and  (\ref{1xy}) are those we would have in the frame co-moving with Plate~2;  hence we must put $\beta\rightarrow -\beta$, $\omega\rightarrow\omega'$ and $u\rightarrow u'$ everywhere except in  the reflection coefficients. Noting that $w$ is invariant under $\omega\rightarrow\omega'$, $u\rightarrow u'$ and defining the transformed value of $s$ by
\begin{equation} 
s'=\sqrt{u'^2+v^2},
\end{equation}
we can now write down $\sigma'_{2xx}(T_2)$,  $\sigma'_{2xy}(T_2)$ and $S'_{2x}(T_2)$:
\begin{align}
\sigma'_{2xx}&(T_2)  \nonumber \\
=&\frac{\hbar}{16\pi^3}\int^\infty_{-\infty}du'\int^\infty_{-\infty}dv\int^{cs'}_{0} d\omega'\,\coth \left(\frac{\hbar\omega'}{2k_BT_2}\right)e^{-2a|w|}|w|\left[(cs'^2+u'\beta\omega')^2+v^2\beta^2c^2w^2\right]  \nonumber \\[5pt]
&\times\left\{4\,\text{Im}(r_{E2})\frac{\text{Re}(r_{E1})|a_{BB}|^2(cs'^2+u'\beta\omega')^2+\text{Re}(r_{B1})|a_{EB}|^2v^2\beta^2c^2w^2}{\left|(cs'^2+u'\beta\omega)^2a_{EE}a_{BB}+a_{EB}a_{BE}v^2\beta^2c^2w^2\right|^2}+( {\scriptstyle E}\leftrightarrow  {\scriptstyle B})\right\}  \nonumber \\[5pt]
&-\frac{\hbar}{16\pi^3}\int^\infty_{-\infty}du'\int^\infty_{-\infty}dv\int^{\infty}_{cs'} d\omega'\,\coth \left(\frac{\hbar\omega'}{2k_BT_2}\right)w\left[(cs'^2+u'\beta\omega')^2+v^2\beta^2c^2w^2\right]  \nonumber \\[5pt]
&\times\left\{\left(1-\left|r_{E2}\right|^2\right)\frac{\left(1+\left|r_{E1}\right|^2\right)|a_{BB}|^2(cs'^2+u'\beta\omega')^2+\left(1+\left|r_{B1}\right|^2\right)|a_{EB}|^2v^2\beta^2c^2w^2}{\left|(cs'^2+u'\beta\omega')^2a_{EE}a_{BB}+a_{EB}a_{BE}v^2\beta^2c^2w^2\right|^2}\right. \nonumber  \\[5pt]
&\ \ \ \ \ \ \ +({\scriptstyle E}\leftrightarrow {\scriptstyle B})\Bigg\},  \label{2xx} 
\end{align}
\begin{align}
\sigma'_{2xy}&(T_2)  \nonumber \\
=&\frac{\hbar}{16\pi^3}\int^\infty_{-\infty}du'\int^\infty_{-\infty}dv\int^{cs'}_{0} d\omega'\,\coth \left(\frac{\hbar\omega'}{2k_BT_2}\right)e^{-2a|w|}u'\left[(cs'^2+u'\beta\omega')^2+v^2\beta^2c^2w^2\right]  \nonumber \\[5pt]
&\times\left\{4\,\text{Im}(r_{E2})\frac{\text{Im}(r_{E1})|a_{BB}|^2(cs'^2+u'\beta\omega')^2+\text{Im}(r_{B1})|a_{EB}|^2v^2\beta^2c^2w^2}{\left|(cs'^2+u'\beta\omega')^2a_{EE}a_{BB}+a_{EB}a_{BE}v^2\beta^2c^2w^2\right|^2}+( {\scriptstyle E}\leftrightarrow  {\scriptstyle B})\right\}  \nonumber \\[5pt]
&+\frac{\hbar}{16\pi^3}\int^\infty_{-\infty}du'\int^\infty_{-\infty}dv\int^{\infty}_{cs'} d\omega'\,\coth \left(\frac{\hbar\omega'}{2k_BT_2}\right)u'\left[(cs'^2+u'\beta\omega')^2+v^2\beta^2c^2w^2\right]  \nonumber \\[5pt]
&\times\left\{\left(1-\left|r_{E2}\right|^2\right)\frac{\left(1-\left|r_{E1}\right|^2\right)|a_{BB}|^2(cs'^2+u'\beta\omega')^2+\left(1-\left|r_{B1}\right|^2\right)|a_{EB}|^2v^2\beta^2c^2w^2}{\left|(cs'^2+u'\beta\omega')^2a_{EE}a_{BB}+a_{EB}a_{BE}v^2\beta^2c^2w^2\right|^2}\right. \nonumber  \\[5pt]
&\ \ \ \ \ \ \ +({\scriptstyle E}\leftrightarrow {\scriptstyle B})\Bigg\},  \label{2xy} 
\end{align}
\begin{align}
S'_{2x}&(T_2)  \nonumber \\
=&-\frac{\hbar}{16\pi^3}\int^\infty_{-\infty}du'\int^\infty_{-\infty}dv\int^{cs'}_{0} d\omega'\,\coth \left(\frac{\hbar\omega'}{2k_BT_2}\right)e^{-2a|w|}\omega'\left[(cs'^2+u'\beta\omega')^2+v^2\beta^2c^2w^2\right]  \nonumber \\[5pt]
&\times\left\{4\,\text{Im}(r_{E2})\frac{\text{Im}(r_{E1})|a_{BB}|^2(cs'^2+u'\beta\omega')^2+\text{Im}(r_{B1})|a_{EB}|^2v^2\beta^2c^2w^2}{\left|(cs'^2+u'\beta\omega')^2a_{EE}a_{BB}+a_{EB}a_{BE}v^2\beta^2c^2w^2\right|^2}+( {\scriptstyle E}\leftrightarrow  {\scriptstyle B})\right\}  \nonumber \\[5pt]
&-\frac{\hbar}{16\pi^3}\int^\infty_{-\infty}du'\int^\infty_{-\infty}dv\int^{\infty}_{cs'} d\omega'\,\coth \left(\frac{\hbar\omega'}{2k_BT_2}\right)\omega'\left[(cs'^2+u'\beta\omega')^2+v^2\beta^2c^2w^2\right]  \nonumber \\[5pt]
&\times\left\{\left(1-\left|r_{E2}\right|^2\right)\frac{\left(1-\left|r_{E1}\right|^2\right)|a_{BB}|^2(cs'^2+u'\beta\omega')^2+\left(1-\left|r_{B1}\right|^2\right)|a_{EB}|^2v^2\beta^2c^2w^2}{\left|(cs'^2+u'\beta\omega')^2a_{EE}a_{BB}+a_{EB}a_{BE}v^2\beta^2c^2w^2\right|^2}\right. \nonumber  \\[5pt]
&\ \ \ \ \ \ \ +({\scriptstyle E}\leftrightarrow {\scriptstyle B})\Bigg\},  \label{S2x} 
\end{align}

The quantities $\omega$ and $u$ occur in the reflection coefficients $r_{E1}$ and $r_{B1}$ in (\ref{2xx})--(\ref{S2x}); we recall that  $r_{E1}$ and $r_{B1}$ are present both explicitly and also throug
h the definitions (\ref{adef1})--(\ref{adef1}). In their appearance in  (\ref{2xx})--(\ref{S2x}), $\omega$ and $u$ are to be taken as functions of the integration variables $\omega'$ and $u'$ in accordance with the inverse of the relations (\ref{lor}):
\begin{equation}
 \omega=\gamma(\omega'+\beta c u'), \quad u=\gamma(u'+\beta\omega'/c).
\end{equation} 

\subsection{Contribution in frame of Fig.~\ref{fig1}}
We must now consider how to use the results of the previous subsection to find  $\sigma_{2xx}(T_2)$ and  $\sigma_{2xy}(T_2)$, the contribution of Plate~2 to the stress tensor components (\ref{xx}) and (\ref{xy}) in the frame of Fig.~\ref{fig1}. Relating tensor components in different inertial frames is straightforwardly done by a Lorentz transformation, but here we are dealing with one of two contributions to a tensor (the electromagnetic energy-momentum tensor) and when a tensor is decomposed into a sum of two terms the individual terms may not be tensors. If one ignores quantum-vacuum effects and considers the purely thermal electromagnetic energy then it is clear that the contribution of each plate to the energy-momentum tensor separately constitutes a tensor, since each contribution gives the full energy-momentum tensor when the  other plate is at zero temperature. In contrast, the quantum-vacuum part of the energy-momentum tensor cannot be separated into contributions from the two plates, each of which could, in some arrangement, constitute the entire energy-momentum tensor and be separately measured. The contribution of Plate~2 to the purely thermal energy-momentum tensor can thus be Lorentz transformed from the co-moving frame to the frame of Fig.~\ref{fig1}; let us first do this and then consider the quantum-vacuum part of the energy-momentum tensor. Separation of quantum-vacuum from thermal effects in the foregoing is achieved by the identity
\begin{equation} \label{coth}
\coth \left(\frac{\hbar\omega}{2k_BT}\right)=\text{sgn}(\omega)+2\,\text{sgn}(\omega)\left[\exp\left(\frac{\hbar|\omega|}{k_BT}\right)-1\right]^{-1},
\end{equation}
where the first term gives the quantum-vacuum part and the second term, containing the Planck spectrum, gives the thermal-radiation part~\cite{LL}. We denote the thermal part  of a quantity by the subscript $\scriptstyle\mathrm{th}$ and proceed to Lorentz transform the thermal part $\sigma'_{\mathrm{th}2xx}(T_2)$ and  $\sigma'_{\mathrm{th}2xy}(T_2)$ of (\ref{2xx}) and (\ref{2xx}) to obtain $\sigma_{\mathrm{th}2xx}(T_2)$ and  $\sigma_{\mathrm{th}2xy}(T_2)$.

 In the gap between the plates, the energy-momentum tensor $T^{\mu\nu}$ is given by~\cite{jac}
\begin{equation}  \label{em}
T^{\mu\nu}=
\left(
\begin{array}{cc}
\frac{1}{2}(\varepsilon_0 E^2+\mu^{-1}_0B^2)  &  \mathbf{S}/c    \\
  \mathbf{S}/c &   -\bm{\sigma}
\end{array}
\right),
\end{equation}
where  $\mathbf{S}$ is the Poynting vector (\ref{poyn}) and $\bm{\sigma}$ is the Maxwell stress tensor (\ref{stress}). (The conventional definition  (\ref{stress}) takes the Maxwell stress tensor as minus the spatial part of the energy-momentum tensor~\cite{jac}.) We have been writing indices in the lower position on $\mathbf{S}$ and $\bm{\sigma}$, which is consistent with (\ref{em}) as we choose a metric signature of $+2$:
\begin{equation}
g_{\mu\nu}=\text{diag}(-1,1,1,1).
\end{equation}
Equation~(\ref{em}) is written without primes and refers to the frame of Fig.~\ref{fig1}. In the frame co-moving with Plate~2 the energy-momentum tensor has the value $T'^{\mu\nu}$, related to $T^{\mu\nu}$ by~\cite{Telephone}
\begin{equation} \label{lorT}
T^{\mu\nu}=\Lambda^\mu_{\ \lambda}\Lambda^\nu_{\ \rho}T'^{\lambda\rho}, \qquad 
\Lambda^\mu_{\ \lambda}=
\left(
\begin{array}{cccc}
 \gamma &  0  &  \gamma \beta & 0 \\
  0 &  1  & 0 & 0 \\
   \gamma \beta &  0 &    \gamma & 0 \\
 0 & 0 & 0 & 1
\end{array}
\right).
\end{equation}
From (\ref{em}) and (\ref{lorT}), we see that  $\sigma'_{\mathrm{th}2xx}(T_2)$ and  $\sigma'_{\mathrm{th}2xy}(T_2)$ are related to $\sigma_{\mathrm{th}2xx}(T_2)$ and  $\sigma_{\mathrm{th}2xy}(T_2)$ by
\begin{equation}  \label{stresstrans}
\sigma_{\mathrm{th}2xx}(T_2)=\sigma'_{\mathrm{th}2xx}(T_2), \qquad \sigma_{2xy}(T_2)=\gamma\left[\sigma'_{\mathrm{th}2xy}(T_2)-\beta S'_{\mathrm{th}x}(T_2)/c\right].  
\end{equation}
Thus no change is required in the thermal part of (\ref{2xx}) to obtain $\sigma_{\mathrm{th}2xx}(T_2)$. In contrast, $\sigma_{\mathrm{th}2xy}(T_2)$ is given by the linear combination (\ref{stresstrans}) of the thermal parts of (\ref{2xy}) and (\ref{S2x}). This linear combination has a very simple effect, however, because the integrands in  (\ref{2xy}) and (\ref{S2x}) differ only by a factor of  $-\frac{\omega'}{u'}$; we see that (\ref{stresstrans}) gives a factor in the integrands of
\begin{equation}
\gamma(u'+\beta\omega'/c)=u.
\end{equation}
Thus the only change required in the thermal part of (\ref{2xy}) to obtain  $\sigma_{\mathrm{th}2xy}(T_2)$ is to replace the overall factor of $u'$ in the integrands by $u$. 

When adding the contributions from each plate to find the stress tensor components (\ref{xx}) and (\ref{xy}) we should use the same dummy integration variables in each contribution. We therefore also transform the integration variables $\omega',u'$ in the contribution of Plate~2 to $\omega,u$ using the fact that the Lorentz transformation (\ref{lor}) has unit Jacobian~\cite{Telephone}. The factor $cs'^2+u'\beta\omega'$ that appears in (\ref{2xx})--(\ref{S2x}) can be written in terms of unprimed quantities through the identity
\begin{equation} \label{termtrans}
cs'^2+u'\beta\omega'=cs^2-u\beta\omega.
\end{equation}
We must also consider the limits on the frequency integral when we transform $\omega'$ in (\ref{2xx}) and (\ref{2xy}) to $\omega$. The Lorentz transformation in question does not change the value of $w$ and the limits $cs'$ in the integrations over $\omega'$, which represents the separation between real and imaginary $w$, become a limit $cs$ in the integrations over $\omega$. From (\ref{lor}), the limit $0$  in the integrations over $\omega'$ corresponds to a limit $\beta cu$ in the integrations over $\omega$. The  integrations $\int_0^{cs'}d\omega'$ over positive values of $\omega'$ in the evanescent-wave parts of  (\ref{2xx}) and (\ref{2xy}) thus turn into integrations $\int_{\beta cu}^{cs}d\omega$ which range over positive and negative values of $\omega$, since $u$ takes values from $-\infty$ to $\infty$. Overall, performing the change in integration variables, (\ref{stresstrans}) and (\ref{termtrans}) give $\sigma_{\mathrm{th}2xx}(T_2)$ and $\sigma_{\mathrm{th}2xy}(T_2)$ as
\begin{align}
\sigma_{\mathrm{th}2xx}&(T_2)
=\frac{\hbar}{16\pi^3}\int^\infty_{-\infty}du\int^\infty_{-\infty}dv\int^{cs}_{\beta cu} d\omega\,2\left[\exp\left(\frac{\hbar\omega'}{k_BT_2}\right)-1\right]^{-1}  \nonumber \\
&\times e^{-2a|w|}|w|\left[(cs^2-u\beta\omega)^2+v^2\beta^2c^2w^2\right]  \nonumber \\[5pt]
&\times\left\{4\,\text{Im}(r_{E2})\frac{\text{Re}(r_{E1})|a_{BB}|^2(cs^2-u\beta\omega)^2+\text{Re}(r_{B1})|a_{EB}|^2v^2\beta^2c^2w^2}{\left|(cs^2-u\beta\omega)^2a_{EE}a_{BB}+a_{EB}a_{BE}v^2\beta^2c^2w^2\right|^2}+( {\scriptstyle E}\leftrightarrow  {\scriptstyle B})\right\}  \nonumber \\[5pt]
&-\frac{\hbar}{16\pi^3}\int^\infty_{-\infty}du\int^\infty_{-\infty}dv\int^{\infty}_{cs} d\omega\,2\left[\exp\left(\frac{\hbar\omega'}{k_BT_2}\right)-1\right]^{-1}w\left[(cs^2-u\beta\omega)^2+v^2\beta^2c^2w^2\right]  \nonumber \\[5pt]
&\times\left\{\left(1-\left|r_{E2}\right|^2\right)\frac{\left(1+\left|r_{E1}\right|^2\right)|a_{BB}|^2(cs^2-u\beta\omega)^2+\left(1+\left|r_{B1}\right|^2\right)|a_{EB}|^2v^2\beta^2c^2w^2}{\left|(cs^2-u\beta\omega)^2a_{EE}a_{BB}+a_{EB}a_{BE}v^2\beta^2c^2w^2\right|^2}\right. \nonumber  \\[5pt]
&\ \ \ \ \ \ \ +({\scriptstyle E}\leftrightarrow {\scriptstyle B})\Bigg\},  \label{2xx0.5} 
\end{align}
\begin{align}
\sigma_{\mathrm{th}2xy}&(T_2)
=\frac{\hbar}{16\pi^3}\int^\infty_{-\infty}du\int^\infty_{-\infty}dv\int^{cs}_{\beta cu} d\omega\,2\left[\exp\left(\frac{\hbar\omega'}{k_BT_2}\right)-1\right]^{-1}  \nonumber \\
&\times e^{-2a|w|}u\left[(cs^2-u\beta\omega)^2+v^2\beta^2c^2w^2\right]  \nonumber \\[5pt]
&\times\left\{4\,\text{Im}(r_{E2})\frac{\text{Im}(r_{E1})|a_{BB}|^2(cs^2-u\beta\omega)^2+\text{Im}(r_{B1})|a_{EB}|^2v^2\beta^2c^2w^2}{\left|(cs^2-u\beta\omega)^2a_{EE}a_{BB}+a_{EB}a_{BE}v^2\beta^2c^2w^2\right|^2}+( {\scriptstyle E}\leftrightarrow  {\scriptstyle B})\right\}  \nonumber \\[5pt]
&+\frac{\hbar}{16\pi^3}\int^\infty_{-\infty}du\int^\infty_{-\infty}dv\int^{\infty}_{cs} d\omega\,2\left[\exp\left(\frac{\hbar\omega'}{k_BT_2}\right)-1\right]^{-1}u\left[(cs^2-u\beta\omega)^2+v^2\beta^2c^2w^2\right]  \nonumber \\[5pt]
&\times\left\{\left(1-\left|r_{E2}\right|^2\right)\frac{\left(1-\left|r_{E1}\right|^2\right)|a_{BB}|^2(cs^2-u\beta\omega)^2+\left(1-\left|r_{B1}\right|^2\right)|a_{EB}|^2v^2\beta^2c^2w^2}{\left|(cs^2-u\beta\omega)^2a_{EE}a_{BB}+a_{EB}a_{BE}v^2\beta^2c^2w^2\right|^2}\right. \nonumber  \\[5pt]
&\ \ \ \ \ \ \ +({\scriptstyle E}\leftrightarrow {\scriptstyle B})\Bigg\}.  \label{2xy0.5} 
\end{align}

We can separate the integration over negative values of $\omega$ in the evanescent-wave parts of  (\ref{2xx0.5}) and (\ref{2xy0.5}) through
\begin{equation}  \label{intsplit}
\int_{-\infty}^\infty du\int_{\beta cu}^{cs}d\omega=\int_{-\infty}^0 du\int_{-\beta c|u|}^{0}d\omega+\int_{-\infty}^0 du\int_{0}^{cs}d\omega+\int^{\infty}_0 du\int_{\beta c|u|}^{cs}d\omega,
\end{equation}
where we assume that $\beta>0$; it is easy to verify that a similar treatment of the case $\beta<0$ gives the same final result (\ref{2xxf}) and  (\ref{2xyf}). Through a variable transformation $\omega\rightarrow -\omega,u\rightarrow -u$ in the first term on the right-hand side of (\ref{intsplit}) we can write the results (\ref{2xx0.5}) and (\ref{2xy0.5}) in a form where all integrations are over positive frequency. To do this we need to consider the effect of $\omega\rightarrow -\omega,u\rightarrow -u$ on the integrands in the evanescent-wave parts. Note first from (\ref{lor}) that this transformation implies $\omega'\rightarrow -\omega'$. We thus need to consider changing the sign of the frequency at which the reflection coefficients of both plates are evaluated. A change in sign of the frequency at which they are evaluated is equivalent to taking the complex conjugate of the reflection coefficients, whereby the real parts are unaltered but the imaginary parts change sign. From this one can see that a simultaneous change in sign of $\omega$, $u$ and $\omega'$ in the integrands of the evanescent-wave parts of (\ref{2xx0.5}) and (\ref{2xy0.5}) produces an overall change in sign of the integrands together with a factor of $-1$ in the exponent of the Planck distribution factor. After the change $\omega\rightarrow -\omega,u\rightarrow -u$, the first and last terms in (\ref{intsplit}) can thus be combined to give an integral $\int^{\infty}_0 du\int_{0}^{cs}d\omega$ whereby (\ref{2xx0.5}) and (\ref{2xy0.5}) can be written
\begin{align}
\sigma_{\mathrm{th}2xx}&(T_2)=\frac{\hbar}{16\pi^3}\int^\infty_{-\infty}du\int^\infty_{-\infty}dv\int^{cs}_{0} d\omega\,2\,\text{sgn}(\omega')\left[\exp\left(\frac{\hbar|\omega'|}{k_BT_2}\right)-1\right]^{-1}  \nonumber \\
&\times e^{-2a|w|}|w|\left[(cs^2-u\beta\omega)^2+v^2\beta^2c^2w^2\right]  \nonumber \\[5pt]
&\times\left\{4\,\text{Im}(r_{E2})\frac{\text{Re}(r_{E1})|a_{BB}|^2(cs^2-u\beta\omega)^2+\text{Re}(r_{B1})|a_{EB}|^2v^2\beta^2c^2w^2}{\left|(cs^2-u\beta\omega)^2a_{EE}a_{BB}+a_{EB}a_{BE}v^2\beta^2c^2w^2\right|^2}+( {\scriptstyle E}\leftrightarrow  {\scriptstyle B})\right\}  \nonumber \\[5pt]
&-\frac{\hbar}{16\pi^3}\int^\infty_{-\infty}du\int^\infty_{-\infty}dv\int^{\infty}_{cs} d\omega\,2\left[\exp\left(\frac{\hbar\omega'}{k_BT_2}\right)-1\right]^{-1}w\left[(cs^2-u\beta\omega)^2+v^2\beta^2c^2w^2\right]  \nonumber \\[5pt]
&\times\left\{\left(1-\left|r_{E2}\right|^2\right)\frac{\left(1+\left|r_{E1}\right|^2\right)|a_{BB}|^2(cs^2-u\beta\omega)^2+\left(1+\left|r_{B1}\right|^2\right)|a_{EB}|^2v^2\beta^2c^2w^2}{\left|(cs^2-u\beta\omega)^2a_{EE}a_{BB}+a_{EB}a_{BE}v^2\beta^2c^2w^2\right|^2}\right. \nonumber  \\[5pt]
&\ \ \ \ \ \ \ +({\scriptstyle E}\leftrightarrow {\scriptstyle B})\Bigg\},  \label{2xxf0} 
\end{align}
\begin{align}
\sigma_{\mathrm{th}2xy}&(T_2)=\frac{\hbar}{16\pi^3}\int^\infty_{-\infty}du\int^\infty_{-\infty}dv\int^{cs}_{0} d\omega\,2\,\text{sgn}(\omega')\left[\exp\left(\frac{\hbar|\omega'|}{k_BT_2}\right)-1\right]^{-1} \nonumber \\
&\times e^{-2a|w|}u\left[(cs^2-u\beta\omega)^2+v^2\beta^2c^2w^2\right]  \nonumber \\[5pt]
&\times\left\{4\,\text{Im}(r_{E2})\frac{\text{Im}(r_{E1})|a_{BB}|^2(cs^2-u\beta\omega)^2+\text{Im}(r_{B1})|a_{EB}|^2v^2\beta^2c^2w^2}{\left|(cs^2-u\beta\omega)^2a_{EE}a_{BB}+a_{EB}a_{BE}v^2\beta^2c^2w^2\right|^2}+( {\scriptstyle E}\leftrightarrow  {\scriptstyle B})\right\}  \nonumber \\[5pt]
&+\frac{\hbar}{16\pi^3}\int^\infty_{-\infty}du\int^\infty_{-\infty}dv\int^{\infty}_{cs} d\omega\,2\left[\exp\left(\frac{\hbar\omega'}{k_BT_2}\right)-1\right]^{-1}u\left[(cs^2-u\beta\omega)^2+v^2\beta^2c^2w^2\right]  \nonumber \\[5pt]
&\times\left\{\left(1-\left|r_{E2}\right|^2\right)\frac{\left(1-\left|r_{E1}\right|^2\right)|a_{BB}|^2(cs^2-u\beta\omega)^2+\left(1-\left|r_{B1}\right|^2\right)|a_{EB}|^2v^2\beta^2c^2w^2}{\left|(cs^2-u\beta\omega)^2a_{EE}a_{BB}+a_{EB}a_{BE}v^2\beta^2c^2w^2\right|^2}\right. \nonumber  \\[5pt]
&\ \ \ \ \ \ \ +({\scriptstyle E}\leftrightarrow {\scriptstyle B})\Bigg\}.  \label{2xyf0} 
\end{align}

The thermal parts of the  components (\ref{xx}) and (\ref{xy}) of the stress tensor between the plates are determined by (\ref{2xxf0}), (\ref{2xyf0}) and the thermal parts of (\ref{1xx}) and (\ref{1xy}). This is the effect of real radiation reflecting back and forth in the gap between the plates, though because of its motion the radiation in the gap arising from Plate~2 does not have a thermal distribution in the frame of Fig.~\ref{fig1}. The plates have a similar effect on the quantum-vacuum radiation in the gap, but such radiation has a featureless ``spectrum" in that all frequencies are present in their zero-point guise. If the real radiation is removed then, in the frame of Fig.~\ref{fig1}, the effect of zero-point radiation is obtained by replacing the spectral distribution of the real radiation from each plate (with an extra factor of $2$) by $\mathrm{sgn}(\omega)$, as is seen from (\ref{coth}). From  (\ref{2xxf0}) and (\ref{2xyf0}) we thus obtain the full contributions in the frame of Fig.~\ref{fig1}, $\sigma_{2xx}(T_2)$ and $\sigma_{2xy}(T_2)$, as

\begin{align}
\sigma_{2xx}&(T_2)=\frac{\hbar}{16\pi^3}\int^\infty_{-\infty}du\int^\infty_{-\infty}dv\int^{cs}_{0} d\omega\left\{1+2\,\text{sgn}(\omega')\left[\exp\left(\frac{\hbar|\omega'|}{k_BT_2}\right)-1\right]^{-1}\right\}  \nonumber \\
&\times e^{-2a|w|}|w|\left[(cs^2-u\beta\omega)^2+v^2\beta^2c^2w^2\right]  \nonumber \\[5pt]
&\times\left\{4\,\text{Im}(r_{E2})\frac{\text{Re}(r_{E1})|a_{BB}|^2(cs^2-u\beta\omega)^2+\text{Re}(r_{B1})|a_{EB}|^2v^2\beta^2c^2w^2}{\left|(cs^2-u\beta\omega)^2a_{EE}a_{BB}+a_{EB}a_{BE}v^2\beta^2c^2w^2\right|^2}+( {\scriptstyle E}\leftrightarrow  {\scriptstyle B})\right\}  \nonumber \\[5pt]
&-\frac{\hbar}{16\pi^3}\int^\infty_{-\infty}du\int^\infty_{-\infty}dv\int^{\infty}_{cs} d\omega\left\{1+2\left[\exp\left(\frac{\hbar\omega'}{k_BT_2}\right)-1\right]^{-1}\right\} \nonumber \\[5pt]
&\times w\left[(cs^2-u\beta\omega)^2+v^2\beta^2c^2w^2\right]  \nonumber \\[5pt]
&\times\left\{\left(1-\left|r_{E2}\right|^2\right)\frac{\left(1+\left|r_{E1}\right|^2\right)|a_{BB}|^2(cs^2-u\beta\omega)^2+\left(1+\left|r_{B1}\right|^2\right)|a_{EB}|^2v^2\beta^2c^2w^2}{\left|(cs^2-u\beta\omega)^2a_{EE}a_{BB}+a_{EB}a_{BE}v^2\beta^2c^2w^2\right|^2}\right. \nonumber  \\[5pt]
&\ \ \ \ \ \ \ +({\scriptstyle E}\leftrightarrow {\scriptstyle B})\Bigg\},  \label{2xxf} 
\end{align}
\begin{align}
\sigma_{2xy}&(T_2)=\frac{\hbar}{16\pi^3}\int^\infty_{-\infty}du\int^\infty_{-\infty}dv\int^{cs}_{0} d\omega\left\{1+2\,\text{sgn}(\omega')\left[\exp\left(\frac{\hbar|\omega'|}{k_BT_2}\right)-1\right]^{-1}\right\} \nonumber \\
&\times e^{-2a|w|}u\left[(cs^2-u\beta\omega)^2+v^2\beta^2c^2w^2\right]  \nonumber \\[5pt]
&\times\left\{4\,\text{Im}(r_{E2})\frac{\text{Im}(r_{E1})|a_{BB}|^2(cs^2-u\beta\omega)^2+\text{Im}(r_{B1})|a_{EB}|^2v^2\beta^2c^2w^2}{\left|(cs^2-u\beta\omega)^2a_{EE}a_{BB}+a_{EB}a_{BE}v^2\beta^2c^2w^2\right|^2}+( {\scriptstyle E}\leftrightarrow  {\scriptstyle B})\right\}  \nonumber \\[5pt]
&+\frac{\hbar}{16\pi^3}\int^\infty_{-\infty}du\int^\infty_{-\infty}dv\int^{\infty}_{cs} d\omega\left\{1+2\left[\exp\left(\frac{\hbar\omega'}{k_BT_2}\right)-1\right]^{-1}\right\}   \nonumber \\[5pt]
&\times u\left[(cs^2-u\beta\omega)^2+v^2\beta^2c^2w^2\right]  \nonumber \\[5pt]
&\times\left\{\left(1-\left|r_{E2}\right|^2\right)\frac{\left(1-\left|r_{E1}\right|^2\right)|a_{BB}|^2(cs^2-u\beta\omega)^2+\left(1-\left|r_{B1}\right|^2\right)|a_{EB}|^2v^2\beta^2c^2w^2}{\left|(cs^2-u\beta\omega)^2a_{EE}a_{BB}+a_{EB}a_{BE}v^2\beta^2c^2w^2\right|^2}\right. \nonumber  \\[5pt]
&\ \ \ \ \ \ \ +({\scriptstyle E}\leftrightarrow {\scriptstyle B})\Bigg\}.  \label{2xyf} 
\end{align}
The components of the stress tensor between the plates that give rise to the Casimir-Lifshitz force are now found from (\ref{xx}) and (\ref{xy}).

\section{The Casimir-Lifshitz force}  \label{CL}
The electromagnetic force on a section of material is given by differences in the electromagnetic stress tensor on the boundaries of that section. To obtain the force we therefore require the stress tensor in the frame in which the medium is at rest.  The results of the previous section were aimed at finding the stress tensor in the rest frame of Plate~1, and without loss of generality we consider the force on this plate. 

The calculations in this paper were performed for infinite half-spaces, as in Lifshitz's pioneering work~\cite{lif55,LL}. Lifshitz's formalism gives the stress tensor due solely to the presence of the cavity, which vanishes when the plates are infinitely far apart ($a\rightarrow\infty$); his results are valid for plates that are thick enough for the cavity boundary to have a negligible effect on the local electromagnetic fields at the external surfaces. A consequence of the different method of calculation used here and in~\cite{ant08} is that contributions to the stress tensor that are independent of the plate separation $a$ are included, even in the case considered by Lifshitz (non-moving plates at the same temperature), and these contributions are of three kinds. Firstly, we have not yet performed the regularization necessary in all calculations of the Casimir-Lifshitz force and so the $a$-independent diverging stress present in the absence of the plates must be removed; for the stress tensor components relevant for the force this divergence occurs only in the propagating-wave contribution to $\sigma_{xx}$ and it will be isolated in the next subsection. Secondly, the formalism used here takes account of the $a$-independent contribution to the stress tensor that each infinite half-space would make in the absence of the other. On its own, each infinite half-space in thermal equilibrium would experience a radiation pressure on its free surface, and this pressure, equal to $\pi^2(k_BT_1)^4/(45\hbar^3c^3)$ for Plate~1, is present in the contribution to $\sigma_{xx}$ as computed here~\cite{ant08}. For non-moving plates ($\beta=0$) at the same temperature ($T_1=T_2$) the (regularized) result for $\sigma_{xx}$ obtained here and in~\cite{ant08} differs from that of Lifshitz by these radiation-pressure terms (see~\cite{ant08} for a detailed discussion). Thirdly, in the situation considered here and in~\cite{ant08}, the heat transfer between the plates causes a momentum transfer that is independent of the plate separation, and this momentum transfer also shows up in the stress tensor~\cite{ant08}. 

If the plates are taken to have a large but finite thickness the total force requires consideration of the stress tensor on the external surfaces, which depends on the details of any thermal radiation in the external region~\cite{ant08}; for example, radiation in the external region that is in thermal equilibrium with the plate will cancel the radiation pressure in the cavity due to that plate. For the moving plates considered here, one may also need to consider lateral motion of the plate relative to external thermal radiation. Our concern here is with the force arising from the stress tensor in the cavity, as this gives the Casimir-Lifshitz force.

Before we proceed to discuss the force on Plate~1 due to the stress tensor in the cavity, let us consider briefly the electromagnetic stress tensor inside the plate. There are always non-vanishing components of the stress tensor inside the plate, even at zero temperature, namely $\sigma_{yy}$ and $\sigma_{zz}$, and the zero-temperature electromagnetic energy density is also non-zero in the material. The non-vanishing zero-temperature stresses and energy density decrease to zero inside the plate as one moves infinitely far from the boundary, since the the local electromagnetic field infinitely far away is unaffected by the boundary.  The components  $\sigma_{yy}$ and $\sigma_{zz}$ do not cause any forces, but the energy density in the plate contributes to the total Casimir-Lifshitz energy, the derivative of which with respect to the plate separation $a$ is minus the perpendicular force. In~\cite{phi09} it was shown that at zero temperature the components of the stress tensor relevant for the forces (the $xx$- and $xy$-components) vanish inside the plate. The presence of thermal radiation, however, produces a non-zero $xy$-component of the stress tensor in the plate as well as in the gap. To see this, note from (\ref{stress}), (\ref{EBdef}), (\ref{freqA}) and
\begin{equation}
\mathbf{\hat{D}}(\mathbf{r},\omega)=\varepsilon_0\varepsilon(\omega)\mathbf{\hat{E}}(\mathbf{r},\omega), \qquad
\mathbf{\hat{B}}(\mathbf{r},\omega)=\mu_0\mu(\omega)\mathbf{\hat{H}}(\mathbf{r},\omega),
\end{equation}
that $\sigma_{xy}$ can be written
\begin{equation}  \label{xyED}
\sigma_{xy}=\langle\hat{E}_y\hat{D}_x+\hat{B}_x\hat{H}_y\rangle,
\end{equation}
including inside the media of the plate. The operator version of the basic continuity properties of electromagnetic fields~\cite{jac} shows that (\ref{xyED}) is continuous across the plate boundary. The presence of a non-vanishing, constant $\sigma_{xy}$ between the plates therefore implies a non-vanishing $\sigma_{xy}$ inside the plate, and in both regions the non-vanishing  $\sigma_{xy}$ is due solely to the thermal radiation.  If Plate~1 has a finite thickness and is not in motion relative to any thermal radiation in the external region, then the component  $\sigma_{xy}$ inside the plate is due solely to incoming thermal radiation from the gap between the plates. The radiation from the gap is absorbed as it moves into Plate~1 and we assume the plate is thick enough for  $\sigma_{xy}$ to be effectively zero on the external surface. The lateral force per unit area on a slice of Plate~1 between $x=x_1$ and $x=x_2$, say, is $\sigma_{xy}(x_1)-\sigma_{xy}(x_2)$ and since $\sigma_{xy}$ is effectively zero on the external surface there is a lateral force per unit area in the $y$-direction given by the (constant) value of  $\sigma_{xy}$ in the gap. The lateral force must of course oppose the relative motion and so act in opposite directions on the two plates. To obtain the force on Plate~2 one should consider the stress tensor in its rest frame, but the fact that the lateral force acts in opposite directions on the two plates can be seen from $\sigma_{xy}$ being constant in the gap since it represents the flow of $y$-momentum in the (positive) $x$-direction~\cite{Telephone}, i.e.\ the direction out of Plate~1 but into Plate~2. 

Consider now the expressions for the perpendicular and lateral forces on Plate~1 due to the electromagnetic stresses in the cavity. Substitution of  (\ref{1xx})  and (\ref{2xxf}) into (\ref{xx}) gives the perpendicular pressure $\sigma_{xx}(T_1,T_2)$; no great simplification results from the addition of (\ref{1xx}) and (\ref{2xx}) so we do not explicitly write the expression for $\sigma_{xx}(T_1,T_2)$. The lateral force per unit area is given by (\ref{xy}), (\ref{1xy})  and (\ref{2xyf}); because of the symmetry of the integrands in (\ref{1xy})  and (\ref{2xyf}) under the interchange $1\leftrightarrow 2$ everywhere except in the first factors, we can write this as 
\begin{align}
\sigma_{xy}&(T_1,T_2)  \nonumber \\
=&\frac{\hbar}{8\pi^3}\int^\infty_{-\infty}du\int^\infty_{-\infty}dv\int^{cs}_{0} d\omega
\left\{ \text{sgn}(\omega')\left[\exp\left(\frac{\hbar|\omega'|}{k_BT_2}\right)-1\right]^{-1}-\left[\exp\left(\frac{\hbar\omega}{k_BT_1}\right)-1\right]^{-1} \right\}  \nonumber \\[5pt]
&\times e^{-2a|w|}u\left[(cs^2-u\beta\omega)^2+v^2\beta^2c^2w^2\right]  \nonumber \\[5pt]
&\times\left\{4\,\text{Im}(r_{E1})\frac{\text{Im}(r_{E2})|a_{BB}|^2(cs^2-u\beta\omega)^2+\text{Im}(r_{B2})|a_{BE}|^2v^2\beta^2c^2w^2}{\left|(cs^2-u\beta\omega)^2a_{EE}a_{BB}+a_{EB}a_{BE}v^2\beta^2c^2w^2\right|^2}+( {\scriptstyle E}\leftrightarrow  {\scriptstyle B})\right\}  \nonumber \\[7pt]
&+\frac{\hbar}{8\pi^3}\int^\infty_{-\infty}du\int^\infty_{-\infty}dv\int^{\infty}_{cs} d\omega\left\{ \left[\exp\left(\frac{\hbar\omega'}{k_BT_2}\right)-1\right]^{-1}-\left[\exp\left(\frac{\hbar\omega}{k_BT_1}\right)-1\right]^{-1} \right\}  \nonumber \\[5pt]
&\times u\left[(cs^2-u\beta\omega)^2+v^2\beta^2c^2w^2\right]  \nonumber \\[5pt]
&\times\left\{\left(1-\left|r_{E1}\right|^2\right)\frac{\left(1-\left|r_{E2}\right|^2\right)|a_{BB}|^2(cs^2-u\beta\omega)^2+\left(1-\left|r_{B2}\right|^2\right)|a_{BE}|^2v^2\beta^2c^2w^2}{\left|(cs^2-u\beta\omega)^2a_{EE}a_{BB}+a_{EB}a_{BE}v^2\beta^2c^2w^2\right|^2}\right. \nonumber  \\[5pt]
&\ \ \ \ \ \ \ +({\scriptstyle E}\leftrightarrow {\scriptstyle B})\Bigg\}.  \label{lat} 
\end{align}
The perpendicular force $\sigma_{xx}(T_1,T_2)$ consists of a quantum-vacuum contribution plus a contribution from the thermal radiation of the plates, as is seen from (\ref{1xx}),  (\ref{coth}) and (\ref{2xxf}). The quantum-vacuum part of $\sigma_{xx}(T_1,T_2)$, which gives the exact perpendicular pressure at zero temperature, has a fairly simple expression~\cite{phi09} that will be derived again in the next subsection. Note that the lateral force (\ref{lat})  is purely an effect of the thermal radiation and vanishes at zero temperature, as shown in~\cite{phi09}. Thermal radiation provides the means of restoring thermodynamic equilibrium, which requires the damping of the relative motion through a lateral force, even when the plates have the same (non-zero) temperature. We see from (\ref{lat}) and that the non-vanishing of the lateral force when the plates have the same (non-zero) temperature is due to the Planck distribution associated with each plate appearing as a function of the co-moving frequency. The vanishing of the lateral force at zero temperature can thus be viewed as a consequence of the Lorentz invariance of quantum zero-point radiation---it has the same ``spectrum" in every inertial frame.

\subsection{The quantum-vacuum contribution}
As shown above, the quantum-vacuum contribution to the Casimir-Lifshitz force is directed perpendicular to the plates and is given by $\sigma_{xx}(0,0)$; from (\ref{xx}), (\ref{1xx}), (\ref{2xxf}) and (\ref{coth}) this can be written
\begin{align}
\sigma_{xx}&(0,0)  \nonumber \\
=&\frac{\hbar}{4\pi^3}\int^\infty_{-\infty}du\int^\infty_{-\infty}dv\int^{cs}_{0} d\omega\,
 e^{-2a|w|}|w|\frac{(cs^2-u\beta\omega)^2+v^2\beta^2c^2w^2} {\left|(cs^2-u\beta\omega)^2a_{EE}a_{BB}+a_{EB}a_{BE}v^2\beta^2c^2w^2\right|^2}  \nonumber \\[5pt]
&\times\left\{\text{Im}(r_{E1}r_{E2})|a_{BB}|^2(cs^2-u\beta\omega)^2+\text{Im}(r_{E1}r_{B2})|a_{BE}|^2v^2\beta^2c^2w^2+( {\scriptstyle E}\leftrightarrow  {\scriptstyle B})\right\}  \nonumber \\[7pt]
&-\frac{\hbar}{8\pi^3}\int^\infty_{-\infty}du\int^\infty_{-\infty}dv\int^{\infty}_{cs} d\omega\,
 w\frac{(cs^2-u\beta\omega)^2+v^2\beta^2c^2w^2} {\left|(cs^2-u\beta\omega)^2a_{EE}a_{BB}+a_{EB}a_{BE}v^2\beta^2c^2w^2\right|^2}  \nonumber \\[5pt]
&\times\left\{\left(1-\left|r_{E1}r_{E2}\right|^2\right)|a_{BB}|^2(cs^2-u\beta\omega)^2+\left(1-\left|r_{E1}r_{B2}\right|^2\right)|a_{BE}|^2v^2\beta^2c^2w^2+({\scriptstyle E}\leftrightarrow {\scriptstyle B})\right\}.  \label{perp0}
\end{align}
Recall that the two (triple) integrals in (\ref{perp0}) are distinguished by $w$ being imaginary in the first but real in the second. This fact, and the overall factor of $w$ in each of the integrals, allows us to write (\ref{perp0}) as the real part of one (triple) integral. We do this by means of the identities
\begin{gather}
\text{Re}\left(e^{2iwa}r_{E1}r_{E2}a^*_{EE}\right)+\frac{1}{2}\left|a_{EE}\right|^2=\frac{1}{2}\left(1-\left|r_{E1}r_{E2}\right|^2\right) \quad \text{($w$ real)},  \label{reid1}  \\
\text{Re}\left(e^{2iwa}r_{E1}r_{B2}a^*_{EB}\right)+\frac{1}{2}\left|a_{EB}\right|^2=\frac{1}{2}\left(1-\left|r_{E1}r_{B2}\right|^2\right)  \quad \text{($w$ real)},  \label{reid2}  \\
\text{Im}\left(e^{2iwa}r_{E1}r_{E2}a^*_{EE}\right)=e^{-2|w|a}\text{Im}\left(r_{E1}r_{E2}\right)  \quad \text{($w$ imaginary)},  \label{imid1}  \\
\text{Im}\left(e^{2iwa}r_{E1}r_{B2}a^*_{EB}\right)=e^{-2|w|a}\text{Im}\left(r_{E1}r_{B2}\right)  \quad \text{($w$ imaginary)},  \label{imid2}
\end{gather}
which show that  (\ref{perp0}) is
\begin{align}
\sigma_{xx}(0,0) =&-\frac{\hbar}{4\pi^3}\text{Re}\int^\infty_{-\infty}du\int^\infty_{-\infty}dv\int^{\infty}_{0} d\omega\,
w\frac{(cs^2-u\beta\omega)^2+v^2\beta^2c^2w^2} {\left|(cs^2-u\beta\omega)^2a_{EE}a_{BB}+a_{EB}a_{BE}v^2\beta^2c^2w^2\right|^2} \nonumber \\[5pt]
&\times\left\{\left[\left(e^{2iwa}r_{E1}r_{E2}a^*_{EE}+\frac{1}{2}\left|a_{EE}\right|^2\right)|a_{BB}|^2(cs^2-u\beta\omega)^2 \right.\right.  \nonumber  \\[5pt]
&\left. \left. +\left(e^{2iwa}r_{E1}r_{B2}a^*_{EB}+\frac{1}{2}\left|a_{EB}\right|^2\right)|a_{BE}|^2v^2\beta^2c^2w^2\right]+( {\scriptstyle E}\leftrightarrow  {\scriptstyle B})\right\}.   \label{perp03}
\end{align}
By multiplying the numerator in (\ref{perp03}) into the factor in braces, we find terms proportional to $(cs^2-u\beta\omega)^2v^2\beta^2c^2w^2$; these can be re-written using the following (admittedly non-trivial) identity:
\begin{align}
&\left(e^{2iwa}r_{E1}r_{E2}a^*_{EE}+\frac{1}{2}\left|a_{EE}\right|^2\right)|a_{BB}|^2+\left(e^{2iwa}r_{E1}r_{B2}a^*_{EB}+\frac{1}{2}\left|a_{EB}\right|^2\right)|a_{BE}|^2+( {\scriptstyle E}\leftrightarrow  {\scriptstyle B})  \nonumber  \\
&=e^{2iwa}\left(r_{E1}r_{E2}a_{BB}a^*_{EB}a^*_{BE}+r_{E1}r_{B2}a_{BE}a^*_{EE}a^*_{BB}\right)+a_{EE}a_{BB}a^*_{EB}a^*_{BE}+a^*_{EE}a^*_{BB}a_{EB}a_{BE}  \nonumber \\
&\ \ \ +( {\scriptstyle E}\leftrightarrow  {\scriptstyle B}).  \label{longid}
\end{align}
Using (\ref{longid}) in the terms proportional to $(cs^2-u\beta\omega)^2v^2\beta^2c^2w^2$ in (\ref{perp03}), it can be written
\begin{align}
\sigma_{xx}&(0,0)   \nonumber \\
=&-\frac{\hbar}{4\pi^3}\text{Re}\int^\infty_{-\infty}du\int^\infty_{-\infty}dv\int^{\infty}_{0} d\omega\,
\frac{w} {\left|(cs^2-u\beta\omega)^2a_{EE}a_{BB}+a_{EB}a_{BE}v^2\beta^2c^2w^2\right|^2} \nonumber \\[5pt]
&\times\bigg\{e^{2iwa}\left[(cs^2-u\beta\omega)^2(r_{B1}r_{B2}a_{EE}+r_{E1}r_{E2}a_{BB})+ (r_{B1}r_{E2}a_{EB}+r_{E1}r_{B2}a_{BE})v^2\beta^2c^2w^2 \right]  \nonumber  \\[5pt]
&  \times\left[(cs^2-u\beta\omega)^2a^*_{EE}a^*_{BB}+a^*_{EB}a^*_{BE}v^2\beta^2c^2w^2\right]   \nonumber \\
& +\left|(cs^2-u\beta\omega)^2a_{EE}a_{BB}+a_{EB}a_{BE}v^2\beta^2c^2w^2\right|^2  \bigg\},   \label{perp04}
\end{align}
which immediately reduces to
\begin{align}
\sigma_{xx}&(0,0)   \nonumber \\
=&-\frac{\hbar}{4\pi^3}\text{Re}\int^\infty_{-\infty}du\int^\infty_{-\infty}dv\int^{\infty}_{0} d\omega\,
\frac{e^{2iwa}w} {(cs^2-u\beta\omega)^2a_{EE}a_{BB}+a_{EB}a_{BE}v^2\beta^2c^2w^2} \nonumber \\[5pt]
&\times \left[(cs^2-u\beta\omega)^2(r_{B1}r_{B2}a_{EE}+r_{E1}r_{E2}a_{BB})+ (r_{B1}r_{E2}a_{EB}+r_{E1}r_{B2}a_{BE})v^2\beta^2c^2w^2 \right] \nonumber \\
&-\frac{\hbar}{4\pi^3}\text{Re}\int^\infty_{-\infty}du\int^\infty_{-\infty}dv\int^{\infty}_{0} d\omega\,w. \label{perp05}
\end{align}
The second integral in (\ref{perp05}) is the diverging zero-point stress in the absence of the plates and must be dropped to obtain the Casimir-Lifshitz force~\cite{LL}. The quantum-vacuum force is thus given by the first integral in (\ref{perp05}), which reproduces the result of~\cite{phi09}. As in~\cite{phi09}, we can write (\ref{perp05}) as an an integral over imaginary frequencies, where the integrand is better behaved. The permittivities and permeabilities are analytic on the upper-half complex-$\omega$ plane~\cite{LLcm} and $w$ acquires a positive imaginary part in this region, so the integrand in the force term in (\ref{perp05}) is also analytic on the upper-half complex-$\omega$ plane. The $\omega$ integration contour from $0$ to $\infty$ can therefore be stretched into the upper-half complex plane so that it runs from $0$ to $i\infty$ and then back to $\infty$ on the real axis in a quarter circle. Since the integrand has no contribution from infinite frequency (where the materials are transparent), the quarter circle in the contour makes no contribution and the integration can be written from $0$ to $i\infty$. On the positive imaginary frequency axis  the permittivities and permeabilities are real~\cite{LLcm} and $w$ is (positive) imaginary; the integrand is therefore real and we can remove the Re in  (\ref{perp05}). Introducing the notation
\begin{gather}
\frac{\omega}{c}=i\kappa,  \\
A_{EE}=r_{E1}^{-1}r_{E2}^{-1}e^{2a|w|}-1, \qquad A_{BB}=r_{B1}^{-1}r_{B2}^{-1}e^{2a|w|}-1, \\
A_{EB}=r_{E1}^{-1}r_{B2}^{-1}e^{2a|w|}-1, \qquad A_{BE}=r_{B1}^{-1}r_{E2}^{-1}e^{2a|w|}-1,
\end{gather}
the force term in  (\ref{perp05}) then takes the form obtained in~\cite{phi09}:
\begin{align}
&\sigma_{xx}(0,0)   \nonumber \\
&=\frac{\hbar c}{4\pi^3}\int_0^\infty\!\!\! d\kappa\int_{-\infty}^\infty\!\!\! d u\int_{-\infty}^\infty \!\!\!d v\,|w|   \left[\frac{(A_{EE}+A_{BB})(s^2-i \kappa u\beta)^2-(A_{EB}+A_{BE})w^2v^2\beta^2}{A_{EE}A_{BB}(s^2-i \kappa u\beta)^2-A_{EB}A_{BE}w^2v^2\beta^2}\right].   \label{force}
\end{align}

\section{Conclusions}
We have obtained the exact solution for the Casimir-Lifshitz force on parallel plates with arbitrary constant lateral motion, where the plates are at different temperatures and each has arbitrary electric permittivity and magnetic permeability. This generalizes the zero-temperature solution~\cite{phi09} and the solution for non-moving parallel plates at different temperatures~\cite{ant08}. The perpendicular force has contributions from the quantum vacuum and from the thermal radiation. There is also a lateral component of the Casimir-Lifshitz force unless both plates are at zero temperature. The formula for the purely thermal lateral force indicates that the non-existence of a quantum-vacuum friction between parallel plates is to be attributed to the Lorentz invariance of quantum-vacuum energy. 

\acknowledgments
This research is supported by the Royal Society of Edinburgh, the Scottish Government, the Leverhulme Trust, and the Royal Society.

\appendix
\section{Expanded form of the Green tensor}
The Green tensor given by (\ref{Gsol})--(\ref{nsf}) contains the inverse matrix
\begin{equation} \label{mat1}
\left(\mathds{1}-e^{2iwa}\mathbf{R}_1\mathbf{R}_2\right)^{-1} 
\end{equation}
which must be expanded before computing the stress tensor. The method for doing this, which includes several consistency checks, is detailed in~\cite{phi09}. The inverse matrix (\ref{mat1}) was expanded in~\cite{phi09} using a basis constructed from the vectors $\mathbf{n}_{E2}$ and $\mathbf{n}_{B2}$; here it is more convenient to use a basis constructed from $\mathbf{n}_{E1}$ and $\mathbf{n}^-_{B1}$, as follows:
\begin{gather}
  \left(\mathds{1}-e^{2iwa}\mathbf{R}_1\mathbf{R}_2\right)^{-1}=\mathds{1}+\mathbf{M}, \label{inv1} \\
 \mathbf{M}=c_{EE}\,\mathbf{n}_{E1}\otimes\mathbf{n}_{E1}+c_{BB}\,\mathbf{n}^+_{B1}\otimes\mathbf{n}^+_{B1}+c_{EB}\,\mathbf{n}_{E}\otimes\mathbf{n}^+_{B1}+c_{BE}\,\mathbf{n}^+_{B1}\otimes\mathbf{n}_{E1}, \label{inv2}
\end{gather}
The unknown $c$-coefficients in (\ref{inv2}) are found by taking successive dot products of (\ref{mat1}) with polarization vectors; for example, $c_{EB}$ is given by
\begin{equation} \label{dots}
c_{EB}=\left[\left(\mathds{1}-e^{2iwa}\mathbf{R}_1\mathbf{R}_2\right)^{-1}-\mathds{1}\right]_{ab}(\mathbf{n}^+_{B1})_b(\mathbf{n}_{E1})_a.
\end{equation}
Defining $\lambda$, $\nu$ and $\rho$ by
\begin{gather}
 \lambda=\mathbf{n}_{E1}\cdot\mathbf{n}^+_{E2}=\frac{cs^2-u\beta\omega}{s\sqrt{c^2s^2-2cu\beta\omega+(\omega^2-c^2v^2)\beta^2}},  \\
\nu=\mathbf{n}^+_{B1}\cdot\mathbf{n}^+_{E2}=\mathbf{n}_{E1}\cdot\mathbf{n}^+_{B2}=\sqrt{1-\lambda^2},  \\
 \rho=\left[1+e^{4iwa}r_{E1}r_{E2}r_{B1}r_{B2}-e^{2iwa}(r_{E1}r_{E2}\lambda^2+r_{B1}r_{B2}\lambda^2+r_{E2}r_{B1}\nu^2+r_{E1}r_{B2}\nu^2)\right]^{-1}
\end{gather}
we write the results for the $c$-coefficients:
\begin{gather}
c_{EE}=\rho\, e^{2iwa}r_{E1}\left(-e^{2iwa}r_{E2}r_{B1}r_{B2}+r_{E2}\lambda^2+r_{B2}\nu^2\right),  \\
 c_{BB}=\rho\, e^{2iwa}r_{B1}\left(-e^{2iwa}r_{B2}r_{E1}r_{E2}+r_{B2}\lambda^2+r_{E2}\nu^2\right), \\
  c_{BE}=\rho\, e^{2iwa}r_{B1}(r_{E2}-r_{B2})\lambda\nu, \qquad
c_{EB}=\rho\, e^{2iwa}r_{E1}(r_{E2}-r_{B2})\lambda\nu.
\end{gather}

From (\ref{Gsol}) we see that the Green tensor also contains
\begin{equation} \label{mat2}
\mathbf{R}_2\left(\mathds{1}-e^{2iwa}\mathbf{R}_1\mathbf{R}_2\right)^{-1}.
\end{equation}
Using the above results for (\ref{mat1}), this can be written in the form
\begin{align}
\mathbf{R}_2&\left(\mathds{1}-e^{2iwa}\mathbf{R}_1\mathbf{R}_2\right)^{-1}  \nonumber \\
 &=d_{EE}\,\mathbf{n}_{E1}\otimes\mathbf{n}_{E1}+d_{BB}\,\mathbf{n}^+_{B1}\otimes\mathbf{n}^+_{B1}+d_{EB}\,\mathbf{n}_{E1}\otimes\mathbf{n}^+_{B1}+d_{BE}\,\mathbf{n}^+_{B1}\otimes\mathbf{n}_{E1},  \label{inv3}
\end{align}
where the $d$-coefficients are
\begin{gather}
d_{EE}=e^{-2iwa}c_{EE}/r_{E1},  \qquad d_{BB}=e^{-2iwa}c_{BB}/r_{B1},  \\
d_{BE}=e^{-2iwa}c_{BE}/r_{B1},  \qquad d_{EB}=e^{-2iwa}c_{EB}/r_{E1}.
\end{gather}

We insert (\ref{inv1}) and  (\ref{inv3}) in (\ref{Gsol}); using the fact that the vector $\mathbf{k}^+_{1}$ in $\boldsymbol{\mathcal{G}_+}$ (see (\ref{Gpm})) is orthogonal to  $\mathbf{n}_{E1}$ and  $\mathbf{n}_{Bm1}$ in  $\mathbf{T}$ (see (\ref{Tdef})), we obtain
\begin{equation} \label{Gsim}
\mathbf{\widetilde{G}}(x,x'',u,v,\omega) =\left[e^{iwx-iw_1x'}(\mathbf{T}+\mathbf{M}\mathbf{T})+e^{-iw(x-2a)-iw_1x'}\mathbf{N}\mathbf{T}\right]\frac{\mu_1}{2w_1}.
\end{equation}
The Green tensor in the form (\ref{Gsim}) was used to calculate the stress tensor and Poynting vector, as described in the main text.

\end{document}